\documentclass[twocolumn]{aastex63}

\newcommand{\lsun}{L_\odot}
\newcommand{\rsun}{R_\odot}
\newcommand{\msun}{M_\odot}
\newcommand{\lacc}{L_{\rm acc}}
\newcommand{\macc}{\dot{M}_{\rm acc}}
\newcommand{\lstar}{L_\star}
\newcommand{\mstar}{M_\star}
\newcommand{\rstar}{R_\star}

\received{}
\revised{}
\accepted{\today}
\submitjournal{ApJ}

\shorttitle{}
\shortauthors{Fiorellino et al.}

\begin{document}

\title{The accretion process in the DQ Tau binary system}
\correspondingauthor{}
\email{eleonora.fiorellino@csfk.org}

\author[0000-0002-5261-6216]{E. Fiorellino}
\affiliation{Konkoly Observatory, Research Centre for Astronomy and Earth Sciences, E\"otv\"os Lor\'and Research Network (ELKH), Konkoly-Thege Mikl\'os \'ut 15-17, 1121 Budapest, Hungary}

\author[0000-0003-4099-1171]{S. Park}
\affiliation{Konkoly Observatory, Research Centre for Astronomy and Earth Sciences, E\"otv\"os Lor\'and Research Network (ELKH), Konkoly-Thege Mikl\'os \'ut 15-17, 1121 Budapest, Hungary}

\author[0000-0001-7157-6275]{\'A. K\'osp\'al}
\affiliation{Konkoly Observatory, Research Centre for Astronomy and Earth Sciences, E\"otv\"os Lor\'and Research Network (ELKH), Konkoly-Thege Mikl\'os \'ut 15-17, 1121 Budapest, Hungary}
\affiliation{Max Planck Institute for Astronomy, K\"onigstuhl 17, 69117 Heidelberg, Germany}
\affiliation{ELTE E\"otv\"os Lor\'and University, Institute of Physics, P\'azm\'any P\'eter s\'et\'any 1/A, 1117 Budapest, Hungary}

\author[0000-0001-6015-646X]{P. \'Abrah\'am}
\affiliation{Konkoly Observatory, Research Centre for Astronomy and Earth Sciences, E\"otv\"os Lor\'and Research Network (ELKH), Konkoly-Thege Mikl\'os \'ut 15-17, 1121 Budapest, Hungary}
\affiliation{ELTE E\"otv\"os Lor\'and University, Institute of Physics, P\'azm\'any P\'eter s\'et\'any 1/A, 1117 Budapest, Hungary}

\begin{abstract}
  Mass accretion from the circumstellar disk onto the protostar is a fundamental process during star formation. Measuring the mass accretion rate is particularly challenging for stars belonging to binary systems, because it is often difficult to discriminate which component is accreting. DQ~Tau is an almost equal-mass spectroscopic binary system where the components orbit each other every 15.8~days. The system is known to display pulsed accretion, i.e., the periodic modulation of the accretion by the components on eccentric orbit. We present multiepoch ESO/VLT X-Shooter observations of DQ~Tau, with the aim to determine which component of this system is the main accreting source. We use the absorption lines in the spectra to determine the radial velocity of the two components, and measure the continuum veiling as a function of wavelength and time. We fit the observed spectra with non-accreting templates to correct for the photospheric and chromospheric contribution. In the corrected spectra we study in details the profiles of the emission lines and calculate mass accretion rates for the system as a function of orbital phase. In accordance with previous findings, we detect elevated accretion close to periastron. We measure that the accretion rate varies between $10^{-8.5}$ and $10^{-7.3}$~$\msun \,$yr$^{-1}$. The emission line profiles suggest that both stars are actively accreting and the dominant accretor is not always the same component, varying in few orbits. 

\end{abstract}

\keywords{Young stellar objects --- Spectroscopic binary stars --- Circumstellar disks --- Stellar accretion --- Spectroscopy}

\section{Introduction} \label{sect:intro}

In the last decades, the star formation process has been investigated and studied in detail for single low-mass stars. 
In the scenario described by the magnetospheric accretion model \citep[][]{har16}, a strong magnetic field of the young star guides the accretion flow, which is responsible for the interplay between the star and the disk. 
Such strong magnetic field truncates the inner edge of the circumstellar disk in a region where the viscous pressure balances the magnetic pressure, called magnetospheric radius ($R_M$). 
This occurs typically at a distance of $R_M \sim 5 - 10 \, R_\star$. 
This single-star accretion scenario has been confirmed by a large amount of observational evidence of classical T~Tauri stars (CTTs) that 
support the physical mechanism of the inner disk truncation \citep[e.g][]{rig12, ant14, alc14, alc17, man17cha, man19, fio21}.
However, the single-star accretion scenario fails to describe close binary systems (with separation below 100~au), for which the presence of the other component affects the dynamics or even the presence of the circumstellar disks, and, as a consequence, the accretion process.

The current binary system accretion scenario has two main predictions \citep[][]{mon07, tof17}. 
First, the orbital motion clears a certain region around the two components, leading to the formation of up to three disks: two circumstellar disks, one around each component, and a circumbinary disk \citep[CBD,][]{art94}, which has been confirmed through observations \citep[e.g.][]{andrews11}. 
Second, due to the dynamics of the binary system, some material from the CBD  periodically forms accretion streams which fuel the circumstellar disks, if any, or directly the forming-stars \citep{art96}.
Theoretical simulations by \citet{mun16} predicts that in equal-mass binary systems, if the orbit is highly eccentric, the main accretor has a mass accretion rate up to $10-20$ times larger than its companion, and the main accretor changes after about 100 periods, always with larger mass accretion rate. This prediction makes highly eccentric, equal-mass binaries the ideal laboratory to study the accretion and verify hydrodinamical simulations. 

The double-lined spectroscopic binary DQ~Tau is the archetype of the equal-mass, highly eccentric, close binary systems. It was discovered by \citet[][]{mathieu1997}. DQ~Tau is composed by two low-mass M0$-$M1 type CTTs that orbit each other with an orbital period of $P = 15.80158 \pm 0.00066$~days and eccentricity of $0.568$ \citep[][]{cze16}. The rotational period of the stars is $T = 3.017 \pm 0.004$~days \citep{kos18}. The most recent and accurate determination of the stellar mass was given by \citet[][]{cze16}, by combining the orbital solution fitted to radial velocity (RV) data and a disk model fitted to ALMA CO observations. 
Adopting a distance of 155\,pc, they obtained $M_{\star,1+2} = 1.21\pm0.26$~$\msun$. 
Here, we adopt the Gaia EDR3 distance of $d=195$~pc \citep[][]{bailerjones21}. 
Since the mass scales linearly with the distance \citep[][]{cze16}, the total mass of the system at 195~pc is $M_{\star,1+2} = 1.52\pm0.33\,\msun$.
Using the mass ratio computed by \citet[][]{cze16}, $M_{2}/M_{1} =0.93 \pm 0.05 $, the masses of the individual components are $M_{\star, 1} = 0.74\,\msun$ and $M_{\star, 2} = 0.78\,\msun$. 
The new distance, together with constraints from the orbital solution of \citet[][]{cze16}, provides a new inclination of 20.7~deg and new orbital major axis of $a = 0.142$~au.

One among the first estimates of the mass accretion rate ($\macc$) of DQ~Tau was obtained by studying the luminosity of the boundary layer, providing $\log \macc = -7.3$ \citep[][]{har95}.
They provided first results on its orbit and light curves, and they interpreted the accretion process of DQ~Tau in the frame of the ``pulsed accretion model'' \citep[][]{art96}, according to which the accretion is highly modulated by the binaries orbital motion, peaking during periastron passages, with 90\% of the total mass accreted between phases $\phi = 0.7 - 1.3$. At those phases, the accretion rate increases on average by a factor of five, in agreement with \citet{mun16} simulations. This interpretation has been confirmed in the following DQ~Tau accretion studies \citep[e.g.][]{sal10, tof17, kos18, muz19}. However, while the on-average results confirm the model predictions, \citet[][]{tof17} found a complex variability from epoch to epoch, suggesting that the material accretes from the inner edge of the CBD in a more complex way than predicted by models. Accretion events near apastron were also observed and seem to be (quasi)periodic in nature.

To better understand the complex accretion process in DQ~Tau system, we computed the mass accretion rate of this source by analysing the emission lines which trace accretion streams among a wide wavelength range, from the ultraviolet (UVB) to the near-infrared (NIR), for 8 epochs with high resolution spectra. In this way, we were able to study the accretion variability, and to discuss which component is accreting the most.

The structure of the paper is as follows: in Sect.~\ref{sect:obs&data} we report the observations and the data reduction, including telluric correction and flux calibration. In Sect.~\ref{sect:analysis} we describe how we computed the radial velocity, the veiling, the extinction, the spectral type, and the disk mass of DQ~Tau. In Sect.~\ref{sect:acc_general} we describe how we subtracted the photospheric and chromospheric contributions to our spectra, focusing on the analysis of the accretion rates. In Sect.~\ref{sect:conclusions} we summarize our results and draw conclusions.

\begin{deluxetable*}{lrcccccccr}
\tablecaption{Observing log\label{tab:obs}}
\tablewidth{0pt}
\tablehead{
\colhead{Name} & \colhead{DATE OBS} & \colhead{HJD}  & \colhead{Seeing}     & \colhead{Airmass} & \colhead{SNR$_{UVB}$} & \colhead{SNR$_{VIS}$} & \colhead{SNR$_{NIR}$} & \colhead{$\phi^{a}$} & \colhead{$\beta$}\\
\colhead{}     &\colhead{}          & \colhead{}     & \colhead{[$\arcsec$]}& \colhead{}        & \colhead{} & \colhead{} & \colhead{} & \colhead{} & \colhead{[km s$^{-1}$]}
}
\decimalcolnumbers
\startdata
Epoch~1       & 2012 Nov 18       & 2456249.29     & 0.70             & 1.46              & 110                      & 184                    & 276 & 0.905 & 7.941 \\ 
Epoch~2       & 2012 Nov 19       & 2456250.14     & 0.64             & 1.59              & 57                       & 113                    & 227 & 0.959 & 7.863 \\  
Epoch~3       & 2012 Dec 31       & 2456292.21     & 0.83             & 1.72              & 54                       & 169                    & 211 & 0.621 & $-14.047$ \\ 
Epoch~4       & 2013 Jan 02       & 2456294.04     & 0.72             & 1.48              & 49                       & 137                    & 241 & 0.737 & $-14.496$ \\ 
Epoch~5       & 2013 Feb 05       & 2456328.04     & 0.81             & 1.37              & 138                      & 192                    & 230 & 0.889 & $-26.896$ \\ 
Epoch~6       & 2013 Feb 08       & 2456331.03     & 2.79             & 1.37              & 51                       & 129                    & 162 & 0.078 & $-27.565$ \\ 
Epoch~7       & 2013 Mar 10        & 2456361.04     & 1.79             & 1.93              & 47                       & 124                    & 142 & 0.977 & $-29.953$ \\ 
Epoch~8       & 2013 Mar 11        & 2456362.03     & 1.54             & 1.67              & 41                       & 111                    & 133 & 0.040 & $-29.869$\\ 
\enddata
\tablecomments{$^a$The phase $\phi$ is calculated by adopting the values from Table~3 in \citet{cze16}. 
}
\end{deluxetable*}

\section{Observations and Data Reduction} \label{sect:obs&data}

Observations were taken at 8 different epochs between 2012 November 18 and 2013 March 11 with the X-Shooter spectrograph on Very Large Telescope (VLT) at ESO's Paranal Observatory in Chile \citep{ver11}.  
X-Shooter covers simultaneously a wide wavelength range from 300~nm to 2480~nm, and the spectra are divided into three arms, the ultraviolet (UVB, $300 - 550$~nm), the visible (VIS, $500 - 1020$~nm), and the near-infrared (NIR, $1000 - 2480$~nm). 
Observations of DQ~Tau 
were performed with the narrow slits of 0.5$\arcsec$, 0.4$\arcsec$, and 0.4$\arcsec$ in the UVB, VIS, and NIR respectively, leading to spectral resolution of $R \sim 9700, 18\,400$, and 11\,600, respectively. 
The exposure time in each epoch 
was 1220~s, 840~s, and 1200~s in the three arms, respectively.

The data are publicly available in the ESO archive\footnote{\url{http://archive.eso.org/wdb/wdb/adp/phase3_main/query}} \footnote{\url{http://archive.eso.org/scienceportal/home}}. We downloaded data reduced with the VLT/X-Shooter pipeline, which consist of extracted, wavelength-calibrated and flux-calibrated 1-dimensional spectra in tabular format following the established standard for ESO science data products. 

We performed the telluric correction of the VIS and NIR bands using the molecfit tool v3.0.3 \citep{kau15}. This included correction for molecular bands as O$_2$ and H$_2$O lines in VIS \citep{erl98, new98}, where strong residuals remained even after telluric correction. 
We computed the signal-to-noise ratio (SNR) for the spectra after discarding the noisy parts where atmospheric transmission is very low between $JHK$ bands. 
The SNR of the spectra depends on the wavelength and on the actual seeing and airmass, being on average higher in the NIR ($\sim 203 $), than in the VIS ($\sim 132$) and in the UVB ($\sim 68$).  
Further details about the observations are reported in Tab.~\ref{tab:obs}, where the seeing, airmass and  SNR mean values for each epoch are listed. Tab.~\ref{tab:obs} shows also the barycentric velocity correction $\beta$, computed by using the software \textsc{barycorrpy} \citep{kanodia2018}. We applied the appropriate correction to each spectrum. 

\begin{figure*}[t]
    \centering
    \includegraphics[width=\textwidth]{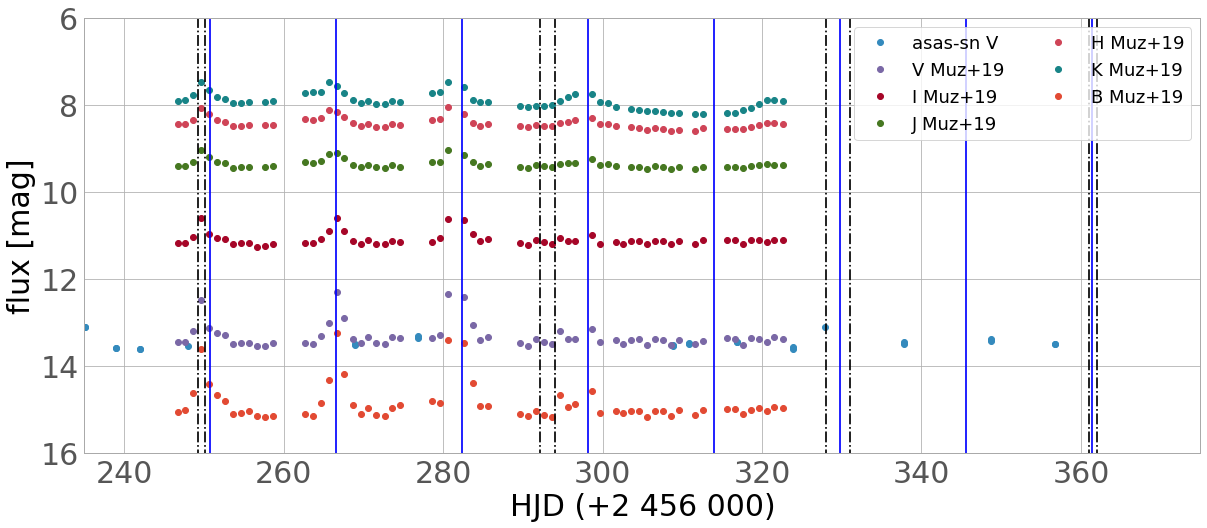}
    \caption{The light curve of DQ~Tau. The epochs of the X-Shooter observations are marked with black point-dashed lines. The epochs of periastron passage are marked with blue lines. Note that Epoch~2 and Epoch~7 correspond to periastron passages. Light curves are obtained from ASAS-SN data or from \citet[][]{muz19} photometry, as described in the legend.}
    \label{fig:ligthcurve_zoom}
\end{figure*}

In order to flux calibrate our spectra, we collected multi-filter photometry for DQ~Tau from the literature \citep[][]{muz19} and data archives \citep[ASAS-SN,][]{sha14, koc17}. The light curves of DQ~Tau are plotted in Fig.~\ref{fig:ligthcurve_zoom}, along with the epochs when the X-Shooter spectra were taken. We used $BVIJHK$ photometry from Figure~\ref{fig:ligthcurve_zoom} for the flux calibration. For the first four epochs, we simply interpolated the available photometry for the exact observing times of the spectra. For Epoch~5, only ASAS-SN $V$-band photometry is available. So we fitted the linear relation between each band and $V$-band photometry. Then we applied this relation to the $V$-band, finding the putative $B$-, $I$-, $J$-, $H$-, and $K$-band photometry. 
Lastly, no photometry in any band is available for Epochs~6, 7, and 8. 
To flux calibrate these spectra, we computed the phase, listed in Tab.~\ref{tab:obs}, of these epochs from orbital parameters provided by \citet[][]{cze16}, and we used the averaged data of the $V$-band for previous periods at the same phase. Then, we applied the fitted relation already used for Epoch~5 to estimate $B$-, $I$-, $J$-, $H$-, and $K$-band photometry. 

We then convert our photometry from magnitudes to flux, by using Bessel zero-fluxes\footnote{\url{http://svo2.cab.inta-csic.es/theory/fps/index.php?mode=browse&gname=Generic&asttype=}}. We linearly interpolated the fluxes for all the wavelength range to flux calibrate the spectra wavelength by wavelength.

The achieved flux calibration of the X-Shooter spectra is shown in Appendix~\ref{app:flux_cal}, in Fig.~\ref{fig:smoothed_spec_with_bandpasses}, epoch by epoch, and in Fig.~\ref{fig:smoothed_spec} for all the epochs.

\section{Analysis}
\label{sect:analysis}
\subsection{Spectral Type and Extinction} \label{sect:spt}
We performed the spectral typing of DQ~Tau by comparing our spectra to a grid of empirical templates from G4 to M9.5 \citep{man13a}, that has a typical step of 1 spectral sub-class for spectral type G and K, and of 0.5 spectral sub-class for M type stars. 
For the comparison, we used only those observed spectra in which the two component were not resolved so that the overall system can be treated as a single star, that means Epoch~3 (see Tab.~\ref{tab:rv}). We reddened the templates, and shift their flux to the one of DQ~Tau in two windows of $\Delta \lambda = 5$~nm, one around $\lambda = 570$~nm, the other around $\lambda = 1230$~nm. By varying the template and the extinction and matching the shape of the overall spectrum and the molecular features of our data with the ones of the templates, we find the best fitting spectral type for the DQ~Tau system is M0, in agreement with the literature \citep[][]{her77, her14, cze16}, with $A_V = 1.7$~mag. After having computed the veiling (see Sect.~\ref{sect:veiling}), we checked this spectral type by correcting the spectra for the veiling. Our best fit is still M0 type with $A_V = 1.7$~mag. This is not surprising, because the veiling is low in the visible, and increases only in the NIR, where we expect the IR-excess (see Fig.~\ref{fig:veil_1}).

\begin{figure}[t]
    \centering
    \includegraphics[width=0.5\textwidth]{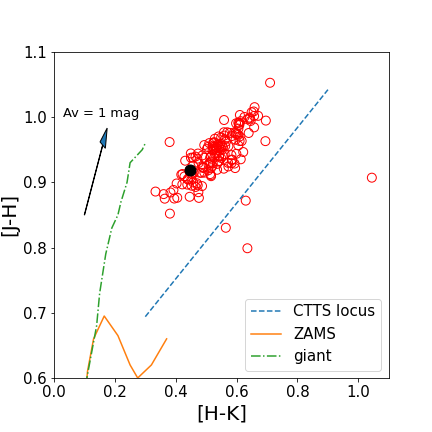}
    \caption{Color diagram of DQ~Tau. Circles are data from \citet{muz19}. The black filled circle corresponds to JHK photometry for Epoch~3. Blue dashed line represent the CTTs locus. 
    The blue arrow shows the shift for a source with 1~mag of extinction. The orange line represents the zero age main sequence, while the green dash-dotted line marks the giant branch from \citet{koo83}.}
    \label{fig:colcol_diag}
\end{figure}

We computed the extinction towards DQ~Tau in an independent way as well, with the help of the color-color diagram shown in Fig.~\ref{fig:colcol_diag}, for Epoch~3, for the same reasons described above. We evaluated the needed extinction to shift the position of the DQ~Tau system on the CTTs locus \citep[][]{mey97}, by using the extinction law by \citet{car89} and $R_V = 3.1$. The resultant extinction we found is $A_V = 1.72 \pm 0.26$~mag, in agreement with the literature \citep[e.g.][]{tof17}, and with our spectral typing.
We adopt this latter value as the interstellar extinction and $A_V = 0$ for the circumstellar extinction, because in \citet[][]{kos18}, the dips due to circumstellar extinction only cause very small changes, and only for 4.85~days in total, which corresponds to 6\% of the time, which suggests all the extinction of DQ~Tau is basically due to the interstellar medium. 

\subsection{Radial Velocity}
\label{sect:rv}

The radial velocity (RV) of DQ~Tau components was measured using SAPHIRES code\footnote{\url{https://github.com/tofflemire/saphires}}, which determines the relative broadening functions (BFs). 
For a complete description of the code, we address the reader to \citet{tof19}. To compute the BFs and RVs, high-resolution empirical templates of non-accreting Class~III spectra were used \citep{man13a}. As a first step, we found the best-fit template spectrum that matches DQ~Tau by determining BFs among the various spectral types ranging from K7 to M3.5. We used only the VIS and NIR spectra, where the continuum and absorption lines are detected with higher SNR than in the UVB. Among the various spectral types in both VIS and NIR, the BF obtained by using M0 template (TYC~7760-283-1) showed the best fit with our spectra. To measure the RVs, we only utilized the VIS, because of their higher resolution and relatively well-corrected telluric absorption features. 
In addition to the best BFs, the M0 template shows the smallest uncertainties of RVs for all observed epochs among all the different spectral types. Consequently, we adopted the M0 template to measure the RVs of DQ~Tau. We note that the M0 template is also compatible with the spectral type we obtained in Sect.~\ref{sect:spt}

The RV of the M0 template spectrum was measured by fitting the BFs with a synthetic spectrum \citep[$T_{\rm eff} = 3750$~K, $\log g = 1$, and solar abundance;][]{coelho2005}. We obtained RV~$ = 6.26 \pm 1.52$~km~s$^{-1}$. The barycentric velocity of the template spectrum was computed by using barycorrpy \citep{kanodia2018} as 2.74~km~s$^{-1}$. 
The template spectrum was therefore shifted by RV and barycentric velocity. The resultant velocity corrected spectrum was used to measure the RVs of DQ Tau. 

\begin{figure}
    \centering
    \includegraphics[width=0.5\textwidth]{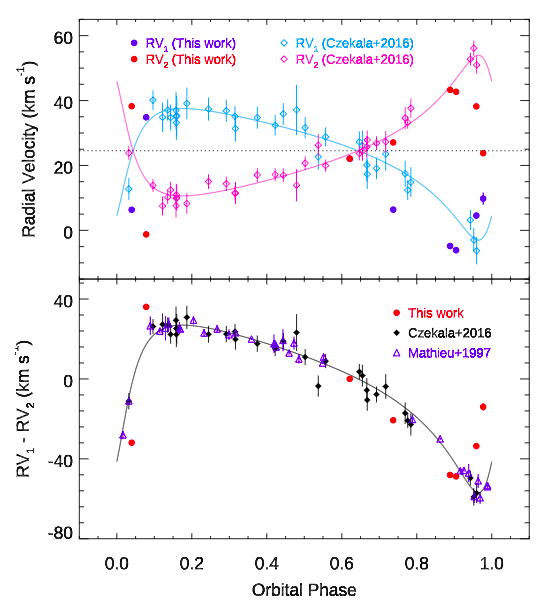}
    \caption{{\it Top}: Radial velocity of DQ~Tau as a function of orbital phase. Blue and red symbols indicate RVs measured in this study (Tab.~\ref{tab:rv}). Sky blue and pink symbols and solid lines present the RVs and orbital solution from \citet{cze16}, respectively.
    {\it Bottom}: Radial velocity differences as a function of orbital phase. Black and red symbols indicate values from \citet{cze16} and this work. Blue symbol represents the data from \citet{mathieu1997}. The uncertainty smaller than the symbol size is not presented.\label{fig:rv}}
\end{figure}

RVs of DQ~Tau were computed by considering 22 fitting regions with about 100~\AA{} wavelength intervals for each epoch. We excluded wavelength ranges where emission lines and strong residuals from telluric absorption lines were present. The mean and standard deviation values are adopted as RV and its uncertainty, respectively. In most epochs, the BFs were double peaked, so we could fit separately the RVs of the two components in all epochs, except for Epoch~3. For Epoch~3, the BF is single-peaked, so the RVs of the two stars must be identical within the uncertainties, so, for this epoch, we give the same value for both stars in Tab.~\ref{tab:rv}, where results are listed. These values are plotted in Fig.~\ref{fig:rv}. 

The top panel in Fig.~\ref{fig:rv} shows the RVs we measured and the ones from \citet{cze16}. To calculate the orbital phase, we adopted the orbital solutions from \citet{cze16}, also shown in the figure as solid lines. The phase coverage of our 8 measurements are not sufficient to determine a Keplerian solution independently, but our results agree with those from \citet[][]{cze16}. There is a slight offset between the absolute values of our RVs and those from \citet{cze16}, but as shown in the bottom panel of Fig.~\ref{fig:rv}, the RV difference of the two components are well consistent with previous studies \citep{cze16, mathieu1997} based on data observed with different telescopes and instruments. 

\begin{deluxetable*}{ccrrccc}
\tabletypesize{\scriptsize} 
\tablecaption{Radial Velocity and Accretion Parameters. \label{tab:rv}}
\tablewidth{0pt}
\tablehead{
\colhead{Epoch} &\colhead{$\phi$} &\colhead{RV$_1$} & \colhead{RV$_2$} & \colhead{$\log (\lacc/ \lsun)$} & \colhead{$\log (\macc/\msun/yr)$} & \colhead{Number of}\\
 &\colhead{} &\colhead{km s$^{-1}$} & \colhead{km s$^{-1}$} &  & &detected lines}
\startdata
 1 & 0.90 & $-6.11 \pm  0.27$ & $42.70 \pm  0.42$ & $-0.144 \pm 0.032$ & $-7.313 \pm 0.032$ & 37\\
 2 & 0.96 &  $4.56 \pm  0.37$ & $38.19 \pm  0.44$ & $-0.138 \pm 0.039$ & $-7.307 \pm 0.039$ & 36\\ 
 3 & 0.62 & $22.08 \pm  0.14$ & $22.08 \pm  0.14$ & $-1.380 \pm 0.034$ & $-8.549 \pm 0.034$ & 20\\
 4 & 0.74 &  $6.38 \pm  0.40$ & $27.08 \pm  0.48$ & $-1.093 \pm 0.032$ & $-8.262 \pm 0.032$ & 25\\
 5 & 0.89 & $-4.82 \pm  0.73$ & $43.28 \pm  0.26$ & $-0.423 \pm 0.031$ & $-7.592 \pm 0.031$ & 37\\
 6 & 0.08 & $34.83 \pm  0.52$ & $-1.24 \pm  0.68$ & $-0.744 \pm 0.035$ & $-7.913 \pm 0.035$ & 28\\
 7 & 0.98 &  $9.79 \pm  1.82$ & $23.78 \pm  0.14$ & $-0.508 \pm 0.024$ & $-7.678 \pm 0.024$ & 34\\
 8 & 0.04 &  $6.33 \pm  0.33$ & $38.25 \pm  0.83$ & $-0.793 \pm 0.046$ & $-7.962 \pm 0.046$ & 34\\
\enddata
\tablecomments{The orbital phase ($\phi$) is computed with respect the nearest previous periastron assuming parameters from \citet{cze16}.}
\end{deluxetable*}

\subsection{Veiling} \label{sect:veiling}
In order to subtract the chromospheric and photospheric contribution from the DQ~Tau spectra, we first evaluated the veiling of the source.
We first normalized the observed spectra ($F_{\nu,\rm obs}$) and the template spectrum ($F_{\nu,\rm temp,0}$), then applied a veiling $V$ to the template:
\begin{equation}
    F_{\nu,\rm temp,veiled} = \frac{F_{\nu,\rm temp,0}+V}{1+V}.
\end{equation}
We call the $V$ parameter {\it continuum veiling} in accordance with the usual practice in the literature \citep[][]{basri1990}, while acknowledging that line veiling may contribute to the measured value. 
It was suggested that emission lines may cause additional veiling (on top of the veiling caused by continuum emission) by \citet[][]{folha1999}. \citet[][]{dodin2012} wrote that narrow emission lines in the post-shock region contribute to the veiling of the photospheric absorption lines and concluded that this contribution is most significant for moderately accreting CTTs. 
Therefore, this could potentially be important for DQ Tau. 
\citet[][]{rei2018} studied high resolution spectra of three T~Tauri stars and found strongly line-dependent veiling: veiling was larger if measured from stronger photospheric lines and lower or absent from weaker lines. They concluded that the best estimate for the true value of the continuum veiling can be obtained by measuring the weakest photospheric lines with equivalent widths (EWs) down to 10~m\AA, and lines with EW above 100~m\AA may already suffer from line veiling.

\begin{figure}
    \centering
    \includegraphics[width=\columnwidth]{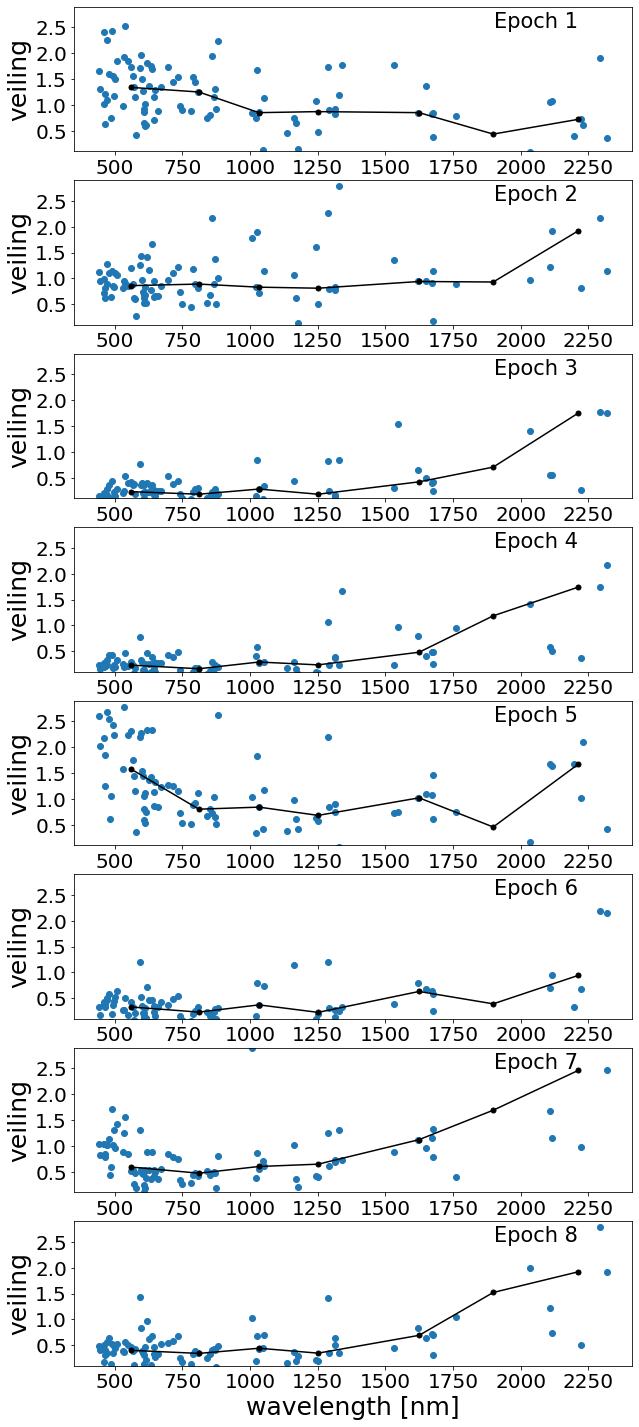}
\caption{Veiling as a function of the wavelength for each epoch. Blue dots are results of the analysis, black dots correspond to median values in 200~nm wide bins.}
    \label{fig:veil_1}
\end{figure}

To find out if line veiling is significant in DQ Tau, we checked the EWs of the photospheric absorption lines we used for our veiling calculation. We found that all of these lines have EW $<$ 100~m\AA in epochs 1, 2, and 5, and only $1-8$ lines have EW $>$ 100~m\AA in the other epochs. 
We found no correlation between the measured veiling and the EWs in any of the epochs. 
Because we used only lines with EW $<$ 100~m\AA for our veiling calculations (with a few exceptions), according to \citet[][]{rei2018}, our result should give the best possible estimate of the true level of the continuum veiling. 
As we see no dependence of the veiling on EW, we can conclude that in the EW range we use for our calculations, line veiling is negligible.
For each epoch, we measured the veiling by studying suitable wavelength ranges around the absorption lines, that are sensitive to the veiling itself. 
We varied the veiling values between 0 and 3, in steps of 0.01, and computed the $\chi^2$ between the veiled template $F_{\nu,temp,veiled}$ and the DQ~Tau spectrum $F_{\nu,obs}$, choosing the veiling value which minimizes the $\chi^2$. 
In the following analysis we use the same veiling for the two components, assuming that in each epoch the primary and the secondary both suffer the same veiling. 
This is not necessarily true, but the limited spectral resolution of X-Shooter, compared to the relatively narrow photospheric absorption lines, and the rather low SNR of the individual absorption lines did not allow us to measure the veiling separately for the two components. Fig.~\ref{fig:veil_1} shows the veiling values measured in 5~nm wide wavelength ranges for selected absorption lines. 
We discarded portions of the spectra where strong residuals of telluric absorption are present. To better outline the general trend how the veiling varies with wavelength and time, we calculated the median of the measured veiling values in 200~nm wide bins. 
We plotted these median values in black in Fig.~\ref{fig:veil_1} which reveals two main trends: ({\it i}) in Epochs~2, 3, 4, 6, 7 and 8, the veiling is almost constant in the UVB and VIS bands, and increasing in the NIR; ({\it ii}) in Epochs~1 and 5, the veiling is constant only in the visible, while it increases both towards shorter and longer wavelengths.

\begin{figure}
    \centering
    \includegraphics[width=0.95\columnwidth]{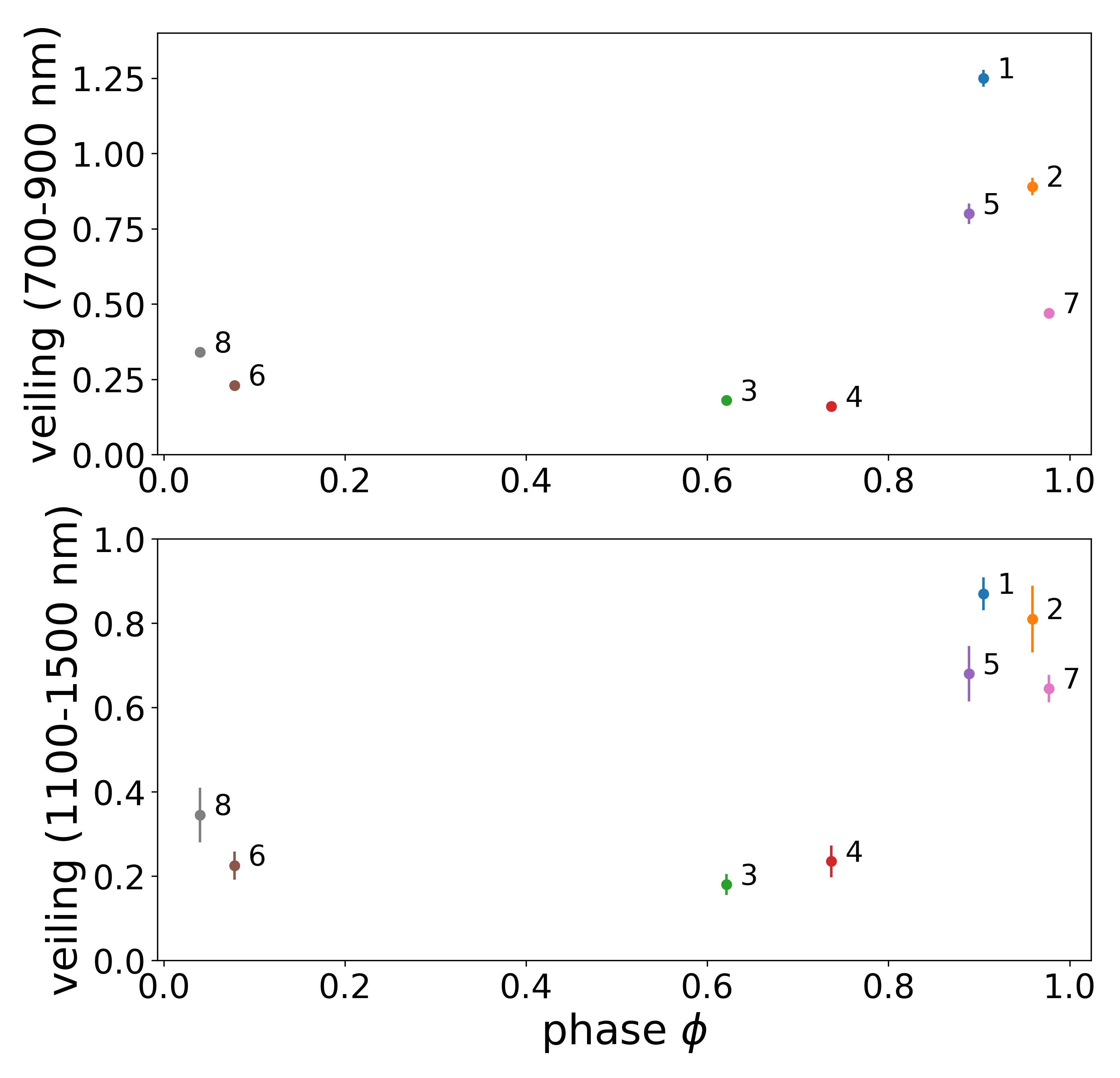}
    \caption{Median veiling values in the optical (top) and near-infrared (bottom) wavelength ranges. Different colors indicate different epochs. Uncertainties smaller than the symbol size are not presented.}
    \label{fig:veiling_vs_phase}
\end{figure}

In Fig.~\ref{fig:veiling_vs_phase}, we show the median veiling as a function of the orbital phase. This figure shows that the veiling varies significantly with the orbital phase, being larger immediately before the periastron both in the NIR and VIS, and with the orbit, being larger for Epochs~1 and 2, which belong to the same orbit, than for other epochs, which belong to a subsequent orbit.

\section{Accretion in DQ~Tau}
\label{sect:acc_general}
DQ~Tau spectra display several emission lines that trace the accretion process. The strength of these lines varies with the orbital phase, and these can be used to calculate the accretion luminosity and the mass accretion rate of the system. In the following, we describe in details the various steps of this procedure, and the relative analysis we performed.

\subsection{Correction for Photospheric and Chromospheric Contributions}
\label{sect:phot_chrom_contributions}

The observed accretion lines of DQ~Tau are contaminated by the photospheric and chromospheric contributions of each component. According to the spectral typing and the radial velocity analysis, the DQ~Tau system is consistently described by two M0 stars. Under the assumption that stars of the same spectral type have similar chromospheric contribution, we fitted both components with the same template, as described below.

For this procedure, we used normalized spectra. The normalization of the spectra was performed considering wavelength ranges of 5~nm. For each wavelength range, we computed the median value of the spectrum, discarding the strong emission lines. We then fitted these median values with a polynomial of second order, and divided the spectra for the fitted line.

For each epoch, we normalized two spectra of the TYC~7760-283-1 star, an M0 Class~III \citep[][]{man13a} also known as V1249~Cen, each shifted to the radial velocity computed in Sect.~\ref{sect:rv}. Then, we summed the two spectra and normalized them again. Hence, we veiled the new template using the values estimated in Sect.~\ref{sect:veiling}. We subtracted the flux of this template from each line of the normalized DQ~Tau spectra which traces the accretion, so that the remaining emission line is completely due to the accretion process.

\subsection{Measuring Emission Line Fluxes}
\label{sect:measurement_acc_li}
After we subtracted the photospheric and chromospheric contributions of the binary components for each epoch, we calculated the fluxes of the accretion tracer emission lines in the following way. First, we fitted a linear curve to the local continuum in a wavelength range of $\Delta \lambda = 2$~nm, centered on the emission line wavelength $\lambda_0$. We modified slightly this range, if needed, taking the one most suitable for each line, for example, avoiding other emission lines, if present, or telluric absorption lines. The line flux was determined by subtracting the local continuum from the spectra and integrating it over the line. 
We computed the noise of the line by multiplying the standard deviation of the local continuum ($RMS$) for the element of wavelength between two pixels $\Delta \lambda$, and multiplying by the square root of the number of pixels within the wavelength range ($N_{pix}$). We considered a line detected when its $SNR > 3$. For those lines that are detected in at least one epoch, we estimated the upper limits in the other epochs as three times the noise:
\begin{equation}
    F_{\rm line}^{upp} = 3 \times \big( \sqrt{N_{pix}} \times RMS \times
    \Delta \lambda \big). 
\end{equation}
Results are shown in Tab.~\ref{tab:flux}.

\begin{table*}[]
 \centering
 \caption{\label{tab:flux} Observed fluxes of accretion tracers after subtracting the photospheric and chromospheric contribution.}
  \resizebox{1.0\textwidth}{!}{%
  \begin{tabular}{lrccccrccc}
\hline
\hline
\colhead{Element} & \colhead{$\lambda$} &\colhead{F$_{\rm line}^1$} & \colhead{F$_{\rm line}^2$} &\colhead{F$_{\rm line}^3$} &\colhead{F$_{\rm line}^4$} &\colhead{F$_{\rm line}^5$} &\colhead{F$_{\rm line}^6$} &\colhead{F$_{\rm line}^7$} &\colhead{F$_{\rm line}^8$}    \\
   &  nm   & $ 10^{-14} \frac{\rm erg}{\rm s \; cm^{2}}$  \\
   &&&&&&&&&\\
\hline
H3(H$\alpha$) & 656.28 & $ 382.554 \pm 0.523 $ & $ 558.038 \pm 0.802 $ & $ 128.549 \pm 0.332 $ & $ 160.253 \pm 0.288 $ & $ 301.567 \pm 0.333 $ & $ 396.943 \pm 0.317 $ & $ 354.495 \pm 0.412 $ & $ 219.908 \pm 0.273 $ \\ 
H4(H$\beta$) & 486.13 & $ 42.697 \pm 0.148 $ & $ 59.401 \pm 0.130 $ & $ 4.981 \pm 0.099 $ & $ 6.686 \pm 0.106 $ & $ 29.826 \pm 0.137 $ & $ 18.946 \pm 0.116 $ & $ 24.077 \pm 0.116 $ & $ 13.795 \pm 0.105 $ \\ 
H5(H$\gamma$) & 434.05 & $ 15.843 \pm 0.445 $ & $ 21.766 \pm 0.150 $ & $ 1.367 \pm 0.059 $ & $ 2.086 \pm 0.065 $ & $ 7.839 \pm 0.196 $ & $ 5.126 \pm 0.080 $ & $ 8.499 \pm 0.087 $ & $ 5.796 \pm 0.077 $ \\ 
H6(H$\delta$) & 410.17 & $ 9.645 \pm 0.103 $ & $ 13.923 \pm 0.068 $ & $ 0.627 \pm 0.030 $ & $ 1.202 \pm 0.034 $ & $ 4.008 \pm 0.044 $ & $ 2.512 \pm 0.036 $ & $ 4.799 \pm 0.041 $ & $ 3.400 \pm 0.036 $ \\ 
H7(H$\epsilon$) & 397.01 & $ 10.882 \pm 0.069 $ & $ 14.144 \pm 0.049 $ & $ 0.855 \pm 0.021 $ & $ 1.486 \pm 0.025 $ & $ 4.988 \pm 0.034 $ & $ 2.741 \pm 0.022 $ & $ 5.065 \pm 0.027 $ & $ 3.483 \pm 0.022 $ \\ 
H8 & 388.90 & $ 4.905 \pm 0.131 $ & $ 7.765 \pm 0.061 $ & $ 0.381 \pm 0.007 $ & $ 0.761 \pm 0.011 $ & $ 1.884 \pm 0.048 $ & $ 1.436 \pm 0.008 $ & $ 2.553 \pm 0.025 $ & $ 1.919 \pm 0.016 $ \\ 
H9 & 383.54 & $ 4.542 \pm 0.197 $ & $ 5.880 \pm 0.103 $ & $ 0.195 \pm 0.007 $ & $ 0.467 \pm 0.009 $ & $ 1.605 \pm 0.085 $ & $ 0.972 \pm 0.014 $ & $ 1.897 \pm 0.038 $ & $ 1.368 \pm 0.021 $ \\ 
H10 & 379.79 & $ 3.479 \pm 0.102 $ & $ 4.395 \pm 0.039 $ & $ 0.124 \pm 0.006 $ & $ 0.312 \pm 0.007 $ & $ 1.236 \pm 0.039 $ & $ 0.674 \pm 0.012 $ & $ 1.439 \pm 0.019 $ & $ 0.923 \pm 0.011 $ \\ 
H11 & 377.06 & $ 2.541 \pm 0.114 $ & $ 3.152 \pm 0.048 $ & $ 0.123 \pm 0.006 $ & $ 0.292 \pm 0.008 $ & $ 0.812 \pm 0.034 $ & $ 0.522 \pm 0.007 $ & $ 1.044 \pm 0.016 $ & $ 0.689 \pm 0.010 $ \\ 
H12 & 375.02 & $ 1.756 \pm 0.101 $ & $ 2.177 \pm 0.048 $ & $ 0.059 \pm 0.004 $ & $ 0.176 \pm 0.005 $ & $ 0.652 \pm 0.043 $ & $ 0.370 \pm 0.006 $ & $ 0.789 \pm 0.019 $ & $ 0.437 \pm 0.012 $ \\ 
H13 & 373.44 & $ 1.863 \pm 0.125 $ & $ 1.651 \pm 0.067 $ & $ 0.039 \pm 0.002 $ & $ 0.109 \pm 0.005 $ & $ 0.573 \pm 0.047 $ & $ 0.273 \pm 0.008 $ & $ 0.629 \pm 0.029 $ & $ 0.291 \pm 0.013 $ \\ 
H14 & 372.19 & $ 1.780 \pm 0.102 $ & $ 1.427 \pm 0.029 $ & $ 0.022 \pm 0.003 $ & $ 0.061 \pm 0.004 $ & $ 0.692 \pm 0.038 $ & $ 0.234 \pm 0.006 $ & $ 0.522 \pm 0.016 $ & $ 0.176 \pm 0.012 $ \\ 
H15 & 371.20 & $ 0.610 \pm 0.151 $ & $ 0.508 \pm 0.110 $ & $ 0.041 \pm 0.004 $ & $ 0.076 \pm 0.009 $ & $ 0.239 \pm 0.059 $ & $ 0.122 \pm 0.019 $ & $ 0.239 \pm 0.043 $ & $ 0.105 \pm 0.021 $ \\ 
Pa5(Pa$\beta$) & 1281.81 & $ 69.714 \pm 0.717 $ & $ 73.728 \pm 0.557 $ & $ 4.930 \pm 0.445 $ & $ 9.130 \pm 0.439 $ & $ 60.821 \pm 0.301 $ & $ 33.397 \pm 0.606 $ & $ 39.672 \pm 0.541 $ & $ 18.321 \pm 0.477 $ \\ 
Pa6(Pa$\gamma$) & 1093.81 & $ 49.409 \pm 0.922 $ & $ 49.751 \pm 2.615 $ & $ 3.597 \pm 0.867 $ & $ 6.103 \pm 0.704 $ & $ 39.439 \pm 1.187 $ & $ 18.951 \pm 1.026 $ & $ 26.125 \pm 0.549 $ & $ 12.532 \pm 0.434 $ \\ 
Pa7(Pa$\delta$) & 1004.94 & $ 33.388 \pm 0.821 $ & $ 32.848 \pm 0.902 $ & $ < 1.261  $ & $ 1.946 \pm 0.462 $ & $ 36.847 \pm 0.839 $ & $ 12.547 \pm 0.538 $ & $ 21.529 \pm 0.753 $ & $ 11.423 \pm 0.779 $ \\ 
Pa9 & 922.90 & $ 20.552 \pm 0.327 $ & $ 12.120 \pm 0.234 $ & $ < 0.683 $ & $ 1.168 \pm 0.167 $ & $ 17.403 \pm 0.277 $ & $ 3.982 \pm 0.209 $ & $ 8.724 \pm 0.221 $ & $ 3.680 \pm 0.171 $ \\ 
Pa10 & 901.49 & $ 19.805 \pm 0.239 $ & $ 7.957 \pm 0.288 $ & $ < 1.359  $ & $ < 0.970 $ & $ 14.578 \pm 0.311 $ & $ 2.409 \pm 0.285 $ & $ 6.022 \pm 0.321 $ & $ 1.933 \pm 0.276 $ \\ 
Br7(Br$\gamma$) & 2166.12 & $ 21.706 \pm 0.586 $ & $ 17.710 \pm 0.593 $ & $ < 1.047 $ & $ 1.217 \pm 0.288 $ & $ 14.483 \pm 0.348 $ & $ 7.181 \pm 0.595 $ & $ 8.520 \pm 0.347 $ & $ 2.905 \pm 0.226 $ \\ 
HeI & 402.62 & $ 0.490 \pm 0.027 $ & $ 0.772 \pm 0.017 $ & $ < 0.045 $ & $ 0.073 \pm 0.017 $ & $ 0.210 \pm 0.013 $ & $ 0.123 \pm 0.014 $ & $ 0.213 \pm 0.013 $ & $ 0.172 \pm 0.012 $ \\ 
HeI & 447.15 & $ 1.635 \pm 0.242 $ & $ 2.336 \pm 0.134 $ & $ < 0.378 $ & $ 0.400 \pm 0.122 $ & $ 0.887 \pm 0.164 $ & $ 0.605 \pm 0.115 $ & $ 0.924 \pm 0.109 $ & $ 0.765 \pm 0.104 $ \\ 
HeI & 471.31 & $ 0.312 \pm 0.045 $ & $ 0.355 \pm 0.053 $ & $ < 0.199 $ & $ < 0.176 $ & $ 0.209 \pm 0.034 $ & $ < 0.158 $ & $ < 0.162 $ & $ 0.154 \pm 0.046 $ \\ 
HeIFeI & 492.19 & $ 7.896 \pm 0.099 $ & $ 2.917 \pm 0.133 $ & $ < 0.457 $ & $ < 0.414 $ & $ 5.675 \pm 0.109 $ & $ < 0.426 $ & $ 1.204 \pm 0.119 $ & $ 0.615 \pm 0.119 $ \\ 
HeI & 501.57 & $ 7.864 \pm 0.114 $ & $ 2.846 \pm 0.128 $ & $ < 0.456 $ & $ < 0.408 $ & $ 5.966 \pm 0.119 $ & $ < 0.402 $ & $ 1.143 \pm 0.119 $ & $ 0.380\pm 0.112 $ \\ 
HeI & 587.56 & $ 7.670 \pm 0.359 $ & $ 9.430 \pm 0.393 $ & $ < 1.649  $ & $ 2.252 \pm 0.566 $ & $ 3.035 \pm 0.425 $ & $ 3.419 \pm 0.457 $ & $ 5.243 \pm 0.346 $ & $ 3.727 \pm 0.446 $ \\ 
HeI & 667.82 & $ 4.419 \pm 0.203 $ & $ 4.465 \pm 0.256 $ & $ < 0.900 $ & $ < 0.845 $ & $ 2.434 \pm 0.163 $ & $ 1.274 \pm 0.241 $ & $ 2.587 \pm 0.262 $ & $ 1.713 \pm 0.236 $ \\ 
HeI & 706.52 & $ 2.732 \pm 0.256 $ & $ 2.677 \pm 0.306 $ & $ < 1.086 $ & $ < 1.020  $ & $ 1.237 \pm 0.205 $ & $ < 0.950  $ & $ 0.944 \pm 0.291 $ & $ < 1.003 $ \\ 
HeII & 468.58 & $ 0.366 \pm 0.054 $ & $ 0.836 \pm 0.053 $ & $ < 0.192  $ & $ < 0.171 $ & $ 0.297 \pm 0.044 $ & $ 0.255 \pm 0.050 $ & $ 0.287 \pm 0.052 $ & $ 0.278 \pm 0.046 $ \\ 
CaII(K) & 393.37 & $ 8.177 \pm 0.158 $ & $ 6.016 \pm 0.064 $ & $ 0.451 \pm 0.010 $ & $ 0.720 \pm 0.013 $ & $ 4.227 \pm 0.073 $ & $ 1.255 \pm 0.012 $ & $ 2.333 \pm 0.032 $ & $ 1.303 \pm 0.015 $ \\ 
CaII(H) & 396.85 & $ 10.67 \pm 0.056 $ & $ 13.810 \pm 0.047 $ & $ 0.876 \pm 0.013 $ & $ 1.452 \pm 0.016 $ & $ 4.989 \pm 0.034 $ & $ 2.733 \pm 0.018 $ & $ 4.984 \pm 0.022 $ & $ 3.374 \pm 0.020 $ \\ 
CaII & 849.80 & $ 46.784 \pm 0.239 $ & $ 12.508 \pm 0.241 $ & $ < 0.676  $ & $ < 0.560 $ & $ 44.201 \pm 0.201 $ & $ 1.393 \pm 0.202 $ & $ 11.62 \pm 0.213 $ & $ 3.794 \pm 0.155 $ \\ 
CaII & 854.21 & $ 42.254 \pm 0.127 $ & $ 15.632 \pm 0.114 $ & $ < 0.562 $ & $ < 0.452 $ & $ 42.988 \pm 0.115 $ & $ < 0.487 $ & $ 11.92 \pm 0.179 $ & $ 0.612 \pm 0.175 $ \\ 
CaII & 866.21 & $ 39.072 \pm 0.221 $ & $ 15.205 \pm 0.210 $ & $ < 1.145 $ & $ < 0.886 $ & $ 40.766 \pm 0.168 $ & $ < 0.844 $ & $ 12.234 \pm 0.321 $ & $ 1.487 \pm 0.273 $ \\ 
NaI & 589.00 & $ 2.162 \pm 0.219 $ & $ 1.157 \pm 0.214 $ & $ < 0.793 $ & $ < 0.821 $ & $ 2.363 \pm 0.270 $ & $ < 0.678 $ & $ < 0.518 $ & $ < 0.701 $ \\ 
NaI & 589.59 & $ 1.260 \pm 0.288 $ & $ < 0.869 $ & $ < 0.860 $ & $ < 0.885$ & $ 1.483 \pm 0.345 $ & $ < 0.684$ & $ < 0.527 $ & $ < 0.664$ \\ 
OI & 777.31 & $ 8.276 \pm 0.275 $ & $ 3.752 \pm 0.297 $ & $ < 1.197 $ & $ < 0.974$ & $ 5.466 \pm 0.225 $ & $ < 0.944$ & $ 2.234 \pm 0.335 $ & $ 1.469 \pm 0.310 $ \\ 
OI & 844.64 & $ 19.353 \pm 0.649 $ & $ 19.197 \pm 0.380 $ & $ < 1.828 $ & $ 2.259 \pm 0.493 $ & $ 17.341 \pm 0.462 $ & $ 7.388 \pm 0.462 $ & $ 8.757 \pm 0.511 $ & $ 4.772 \pm 0.471 $ \\ 
\hline 
\hline
  \end{tabular}  
  }
   \begin{quotation} Reported flux values are not dereddened.
  \end{quotation}  
\end{table*}


\subsection{Accretion luminosity and mass accretion rate}
\label{sect:accretion}

\begin{figure*}[t]
    \centering
    \includegraphics[width=0.95\textwidth]{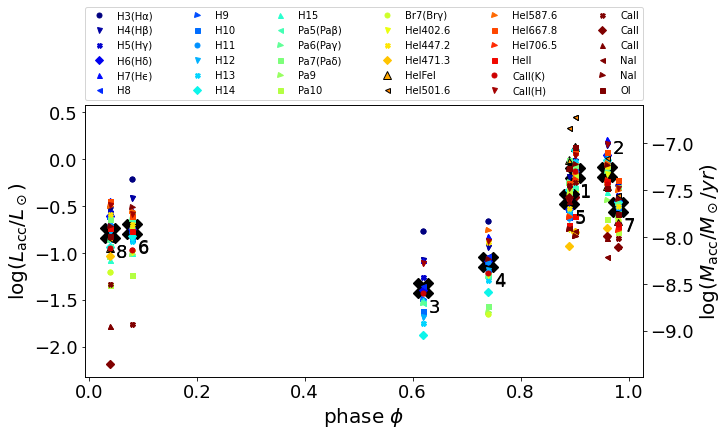}
    \includegraphics[width=0.95\textwidth]{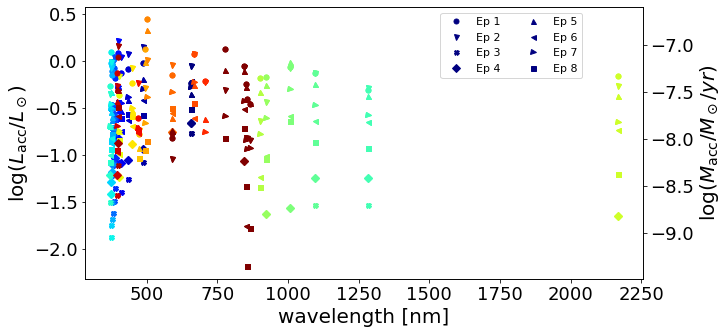}
    \caption{Accretion luminosity and mass accretion rate as a function of the orbital phase (top) and as a function of the wavelength (bottom). Each symbol indicates a different accretion tracer or epoch, as described in the legends.}
    \label{fig:lacc}
\end{figure*}

In order to estimate the accretion luminosity of DQ~Tau, we used the most recent empirical relations between the lines that trace accretion 
($L_{\rm line}$) and the accretion luminosity itself \citep[][]{alc17}:
\begin{equation}
    \log \big( \lacc/\lsun \big) = a_{\rm line} \log \big( L_{\rm line}/ \lsun \big) + b_{\rm line},
    \label{eq:alcala}
\end{equation}
where $a_{\rm line}$ and $b_{\rm line}$ are coefficients that vary with the line, and $L_{\rm line} = 4 \pi d^2 F_{\rm line}$ is the luminosity of the line. The error on $L_{\rm line}$ was computed by using the error propagation formula, considering the error on the distance, and on the line flux. The error on $F_{\rm line}$ was computed in the same way, considering the error on $L_{\rm line}$, and on $a,b$ coefficients. We estimated the accretion luminosity for every line from Tab.~\ref{tab:flux}, and used the mean weighted value of $\lacc$ derived from these lines as the best estimate for $\lacc$ for each epoch. Errors on $\lacc$ are computed by dividing the standard deviation of $\lacc$ computed for every line detected per the square root of the number of used lines. Results are shown in Tab.~\ref{tab:rv}. 

We computed the mass accretion rate $\macc$ using the relation:
\begin{equation}
 \label{eqmacc}
  \macc \sim \left(1 - \frac{\rstar}{R_{\rm in}}\right)^{-1} \frac{\lacc \rstar}{G \mstar},
\end{equation}
where $R_{\rm in}$ is the inner-disk radius which we assume to be $R_{\rm in} \sim 5 R_\star$ \citep{har98}, 
$\mstar = 1.52\,\msun$ and $\rstar = 2.58\,\rsun$ are the mass and radius for the overall system derived by \citet[][]{cze16} and scaled to $d=195\,$pc.
We note that if we use mass and radius of a single star, being the two components of DQ~Tau almost equals ($M_2/ M_1 = 0.93 \pm 0.05$), the factor of two will be in both the numerator and the denominator in Eq.~\ref{eqmacc}, not affecting final results. The error on $\macc$ is computed as for $\lacc$. Being $\macc$ proportional to $\lacc$, the error for each epoch is the same for both $\lacc$ and $\macc$. Indeed, because we are interested in the variability of $\macc$, we have taken into account only the error due to the accretion variability, considering the stellar parameters fixed. Results are listed in Tab.~\ref{tab:rv}. However, the absolute values of the mass accretion rate is affected also by the uncertainties on the stellar parameters, and we will do this when comparing our results to other works. 
In this latter case, the error results to be 0.45~dex.

Both $\lacc$ and $\macc$ are shown as a function of orbital phase in Fig.~\ref{fig:lacc}. 
We see that Epochs~3 and 4, the measurements taken closest to the apastron, show the smallest $\lacc$ and $\macc$. On the contrary, epochs between $\phi = 0.8 - 1.0$ show significantly elevated accretion rates.
Our results that the accretion depends on the orbital phase and is highest nearby periastron support previous observational and numerical results in the literature for DQ~Tau and for eccentric binary systems in general \citep[e.g.][]{mathieu1997, gun02, sal10, dorazio13, farris14, mun16, tof17, kos18, muz19}. 
This is in agreement with the hypothesis that the accretion flow of DQ~Tau can be explained by the ``pulsed accretion model'' \citep[][]{art96}, according to which the accretion is highly modulated by the binaries orbital motion, peaking during periastron passages. 

It is worth considering the system's geometry when interpreting the line flux variations because in theory, these can be caused by rotational modulation as well \citep[e.g.,][]{kurosawa2008,kurosawa2013,romanova2016}. 
The inclination of the rotational axis of the stars in the DQ Tau binary is not known. 
If we assume that the stars' equator is coplanar with the disk, then the DQ~Tau system is viewed nearly pole-on ($i=20.7\,$deg), therefore we do not expect significant rotational modulation. 
For instance, numerical simulations by \citet{kurosawa2008} suggest that the equivalent width of emission lines only change by up to a factor of 1.5 for $i=10\,$deg and up to a factor of 2.5 for high inclination ($i=60-80$\,deg) models, while we observe significantly larger variations in DQ~Tau despite its low inclination. 
Moreover, the line variations observed in DQ~Tau are periodic with the orbital period of the binary and not with the rotational period of the stars. The ratio between the system's orbital period ($15.80158\pm0.00066$\,d) and rotational period ($3.017\pm0.004$\,d) is $5.2375\pm0.0069$, which is far from any possible low-order resonances \citep{kos18}. 
Therefore, we can exclude with high confidence that the axial rotation of the components is synchronized with the orbital motion and changes periodic with the rotational period cannot be confused with changes that are periodic with the binary's orbital period. In conclusion, the observed line flux variations are unlikely to be explained by rotational modulation.

Looking more in detail at Fig.~\ref{fig:lacc}, we note that, when detected, the {He\,{\footnotesize I}} at 501.6~nm and {He\,{\footnotesize I}$+$Fe\,{\footnotesize I} } blend at 492~nm give accretion rates systematically larger than the other lines. On the contrary, the Paschen series usually provide lower accretion rates, especially in Epochs~3, 4, 6, and 8. 
The final errors on the $\log \lacc$ values of individual lines are in the range of 0.17 to 0.73, having a typical uncertainty of 0.33 dex in $\lacc$. 
Fig.~\ref{fig:lacc} shows that the observed differences in accretion luminosity often exceeds these, hinting for a physical reason behind, i.e., that different lines trace different parts of the accretion flow with different physical conditions. 
We will discuss this point in detail in Sect.~\ref{sect:single}.
We note that not all epochs have the same lines detected. To minimize the effect of these uncertainties, we calculated the mean weighted value as the best estimate for the accretion rate.

\begin{figure}
    \centering
    \includegraphics[width=0.5\textwidth]{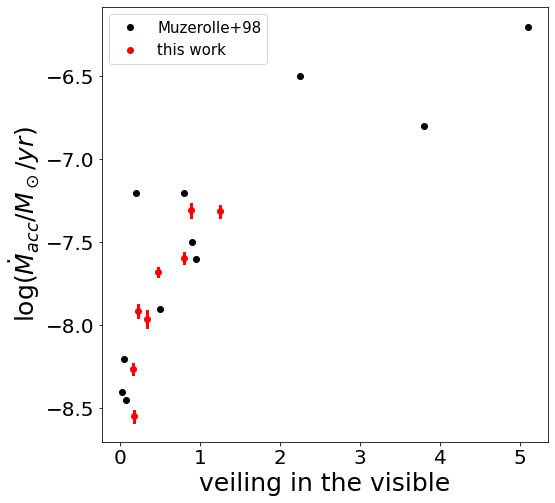}
    \caption{Mass accretion rate as a function of the continuum veiling in the visible.}
    \label{fig:macc_veiling}
\end{figure}

\subsection{Comparing the Continuum Veiling and the Mass Accretion Rate} \label{sect:veil_acc}
The continuum veiling values plotted in Fig.~\ref{fig:veiling_vs_phase} and the accretion rates plotted in Fig.~\ref{fig:lacc} show a similar trend with the orbital phase. Both veiling and accretion rates increase close to periastron. Epochs~1 and 2 display the maximal veiling and maximal accretion rate as well during our monitoring. In the following, we want to compare our results on DQ~Tau with other sources in Taurus, to verify whether this relation between the mass accretion rate and the continuum veiling is peculiar of DQ~Tau or general for the Taurus star forming region. 
For this purpose, we plot in Fig.~\ref{fig:macc_veiling} the mass accretion rate as a function of the median veiling in the visible for a sample in Taurus \citep[black filled dots;][rescaled to 195~pc]{muz98tau}, and for DQ~Tau (red filled dots). 
We note that $\log \macc$ of DQ~Tau increases almost linearly with the continuum veiling. 
The figure shows that the data points for DQ~Tau match the general trend outlined in \citet[][]{muz98tau}.

\subsection{Comparing DQ~Tau with Single Accretors}
\label{sect:single_acc}
\begin{figure}[t]
    \centering
    \includegraphics[width=0.5\textwidth]{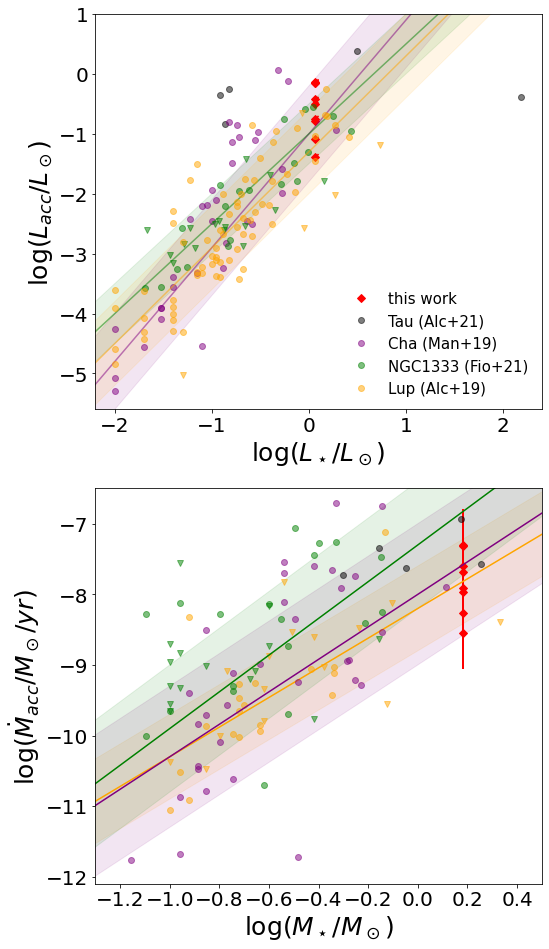}
    \caption{{\it Top}: Accretion luminosity versus stellar luminosity. {\it Bottom}: Mass accretion rate versus stellar mass. In both panels, DQ~Tau results are shown in red, and the Chamaeleon~I, Lupus, NGC~1333, and Taurus CTTs samples \citep[][]{alc19, man19, fio21, alc21} are shown together with their fits \citep[][]{fio21} with the colors described in the legend.}
    \label{fig:acc_tau}
\end{figure}

During periastron, the DQ~Tau components are known to approach each other to within $8 \, \rstar \sim 13 \, \rsun$, which means that their magnetospheres merge \citep[][]{sal10}. 
This may mean that single circumstellar disks, if present, would be disrupted by the magnetosphere during periastron, and, eventually, the accretion columns can generate directly from the CBD to the stars.
Therefore, in terms of the magnetospheric accretion scenario, the DQ~Tau system probably can be considered as a single star when the system is close to (or at) periastron. To verify this, we compared DQ~Tau with the most recent CTTs surveys in Lupus, Chamaeleon~I, NGC1333, and Taurus \citep[][respectively]{alc19,man19,fio21, alc21} in Fig.~\ref{fig:acc_tau}. We plotted DQ~Tau $\lacc$ and $\macc$ total values for the two stars together, at a place that corresponds to the total luminosity and mass of the two stars, respectively. We find that, in general, the accretion rates of DQ~Tau (red filled diamonds) are in agreement with typical ranges of values of single CTTs with the same stellar luminosity and mass. While $\lacc$ perfectly matches CTTs distributions in the $\log \lacc - \log \lstar$ diagram, the $\macc$ of DQ~Tau lies in the lower part of the $\log \macc - \log \mstar$ diagram. In particular, the $\macc$ is compatible with the $\macc$ distribution of $\sim 2$~Myr star-forming regions Chamaeleon~I and Lupus for all the epochs. Thus, our results show that DQ~Tau is accreting with a rate compatible with single stars with similar stellar parameters even when it is accreting less, nearby the apastron. This result seems to confirm the hypothesis that we can study the $\macc$ on DQ~Tau as it was a single star.

\subsection{Accretion Variability over the Binary Orbit}
\begin{figure}[t]
    \centering
    \includegraphics[width=0.5\textwidth]{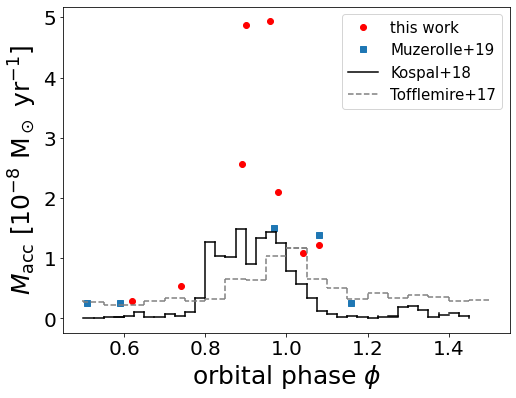}
    
    \caption{Mass accretion rate of DQ~Tau versus phase.} \label{fig:acc_peri}
\end{figure}

To study the accretion variability over the binary orbit, we compare our $\macc$ estimates with results from previous literature, after scaling them to the distance we adopt of 195~pc. 
We plot in Fig.~\ref{fig:acc_peri} the mass accretion rate as a function of the orbital phase. As described in Sect.~\ref{sect:accretion}, we used spectroscopic data and empirical relations between $F_{\rm line}$ and $\lacc$, to compute $\macc$. Our results are shown with red filled circles. The same approach was used by \citet[][]{muz19}, but only for Pa$\beta$ and Br$\gamma$ lines. We computed the mean values of these two $\macc$ estimates, plotting the results as blue filled squares. Differently, \citet[][]{kos18} and \citet[][]{tof17} provided $\macc$ of the DQ~Tau source from photometry data, using an empirical relation between the luminosity excess in the $U-$band and the accretion luminosity \citep[][]{gul98}. Black solid and grey dashed lines are histograms from \citet[][]{kos18} and \citet[][]{tof17}, respectively, where the peaks correspond to the mean value for the same phase of several periods. Fig.~\ref{fig:acc_peri} shows that the trend with respect the phase between the samples is the same, so $\macc$ is larger nearby the periastron in every sample, but results obtained from photometry are about a factor of two smaller than $\macc$ computed from emission lines. This difference can be explained by both the different method used, and the fact that histograms show average values of tens of orbits, while our results and results by \citet[][]{muz19} are computed in a specific phase. The $\macc$ in the four samples were computed by using different techniques, different data at different wavelengths, in different orbits. Moreover, all the methods are based on empirical relations which, therefore, are both affected by large uncertainties. Considering all these differences, it is actually a significant result that the trend shown in this figure is the same for each sample, strengthening the pulsed accretion theory for the DQ~Tau system.

\subsection{Accretion in Binary Systems}

\begin{figure}[t]
    \centering
    \includegraphics[width=0.93\columnwidth]{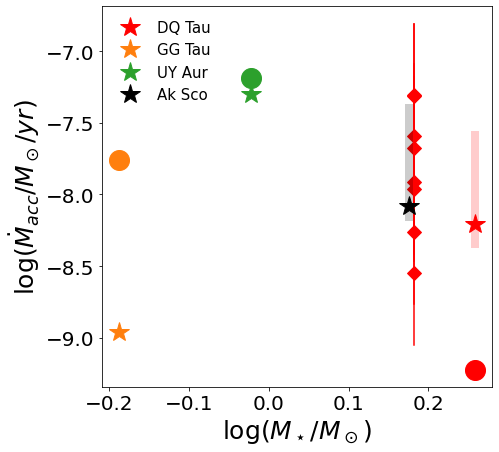}
    
    \caption{Mass accretion rate versus stellar mass. \citet[][]{gut02} simulation results (mean values) for four eccentric binary systems and relative measurement by \citet[][]{gul98} are plotted as stars and circles, respectively. Vertical red and black regions correspond to the accretion variability range obtained from simulations in five different orbits. The colors and names of the systems are reported in the legend. DQ~Tau mass accretion rates of this work are marked with red diamonds.}
    \label{fig:acc_bi}
\end{figure}

An important aspect of accretion in close, eccentric binary systems as DQ~Tau is the absence of single circumstellar disks, and the presence of a CBD around the two stars, from which the material flows to one or both components.
To test this scenario, we compared line observed accretion variability in DQ~Tau with results from theoretical simulations of binary systems provided by \citet[][]{gun02}. Fig.~\ref{fig:acc_bi} shows the mean values of the $\macc$ for four sources. Two of them, GG~Tau and UY~Aur, are wide systems with little eccentricity ($e = 0.25$ and $e = 0.16$, respectively), while DQ~Tau and AK~Sco are close eccentric binaries ($e = 0.556$ and $e=0.469$, respectively). 
The authors conclude that the only reason why $\macc$ of wide binaries is comparable with $\macc$ of close binaries is that the disk mass ($M_{\rm disk}$) for GG~Tau and UY~Aur is three orders of magnitude larger, suggesting that the closer are the two components, the more efficient is the accretion process. 
We scaled the results of DQ~Tau to the distance adopted in this work, for a coherent comparison. 
Fig.~\ref{fig:acc_bi} shows that results by \citet[][]{gun02} are in agreement with the ones we computed nearby the periastron. 
And our estimate for the less accreting epoch, is in agreement within the error with measurements by \citet[][]{gul98}. 
This suggests that \citet[][]{gun02} simulations nicely reproduced the measured mass accretion rate for DQ~Tau at periastron, but somehow overestimated the accretion during other orbital phases. 

\subsection{Accretion from single stars in the DQ~Tau system} \label{sect:single}

\begin{figure*}
    \centering
    \includegraphics[width=0.24\textwidth]{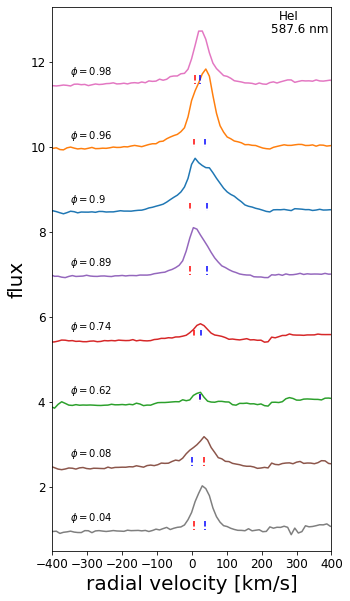}
    \includegraphics[width=0.24\textwidth]{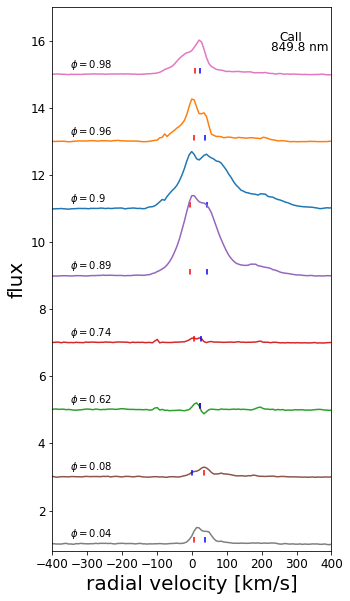}
    \includegraphics[width=0.24\textwidth]{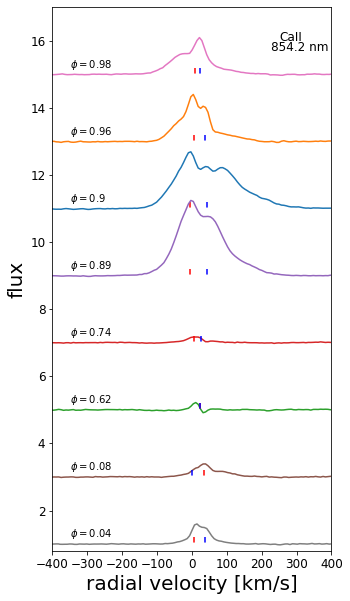}
    \includegraphics[width=0.24\textwidth]{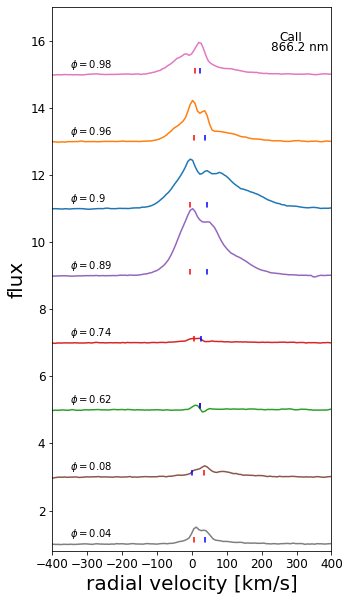}

    \caption{Velocity structure of continuum-normalized {He\,{\footnotesize I}} at $\lambda = 587.65$~nm and {Ca\,{\footnotesize II}} triplet emission lines observed by X-Shooter. Vertical red and blue dashed lines are velocities of the primary and secondary component, respectively. Spectra are ordered from bottom to top by increasing orbital phase, which is labeled adjacent to each spectrum. The line color marks the epoch, using the same colors of Fig.~\ref{fig:lacc}.}
    \label{fig:toff_figs_text}
\end{figure*}

The main goal of this work is to determine whether both the components of DQ~Tau are accreting during each epoch we observed and, if so, which one is accreting the most.  For this purpose, we need two conditions to be satisfied: {\it i}) the two components are spectroscopically resolved; {\it ii}) the accretion luminosity is not too faint, so, the lines which trace accretion are detected. With these general rules in mind, we studied the accretion tracer lines for all the epochs in Fig.~\ref{fig:toff_figs_text} (and in Figs.~\ref{fig:toff_figs}, \ref{fig:toff_figs1}, \ref{fig:toff_figsNIR}, \ref{fig:toff_figsHe}, and \ref{fig:toff_figs_other} in Appendix~\ref{app:lines}), similarly to what was done by \citet[][]{tof19} for TWA~3A system.

\begin{figure}[t]
    \centering
    \includegraphics[width=\columnwidth]{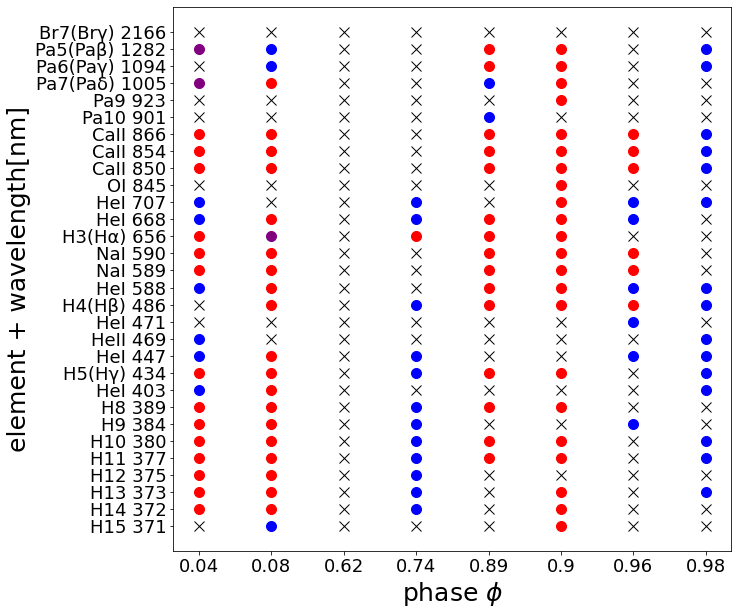}
    \begin{flushright}
    \includegraphics[width=0.91\columnwidth]{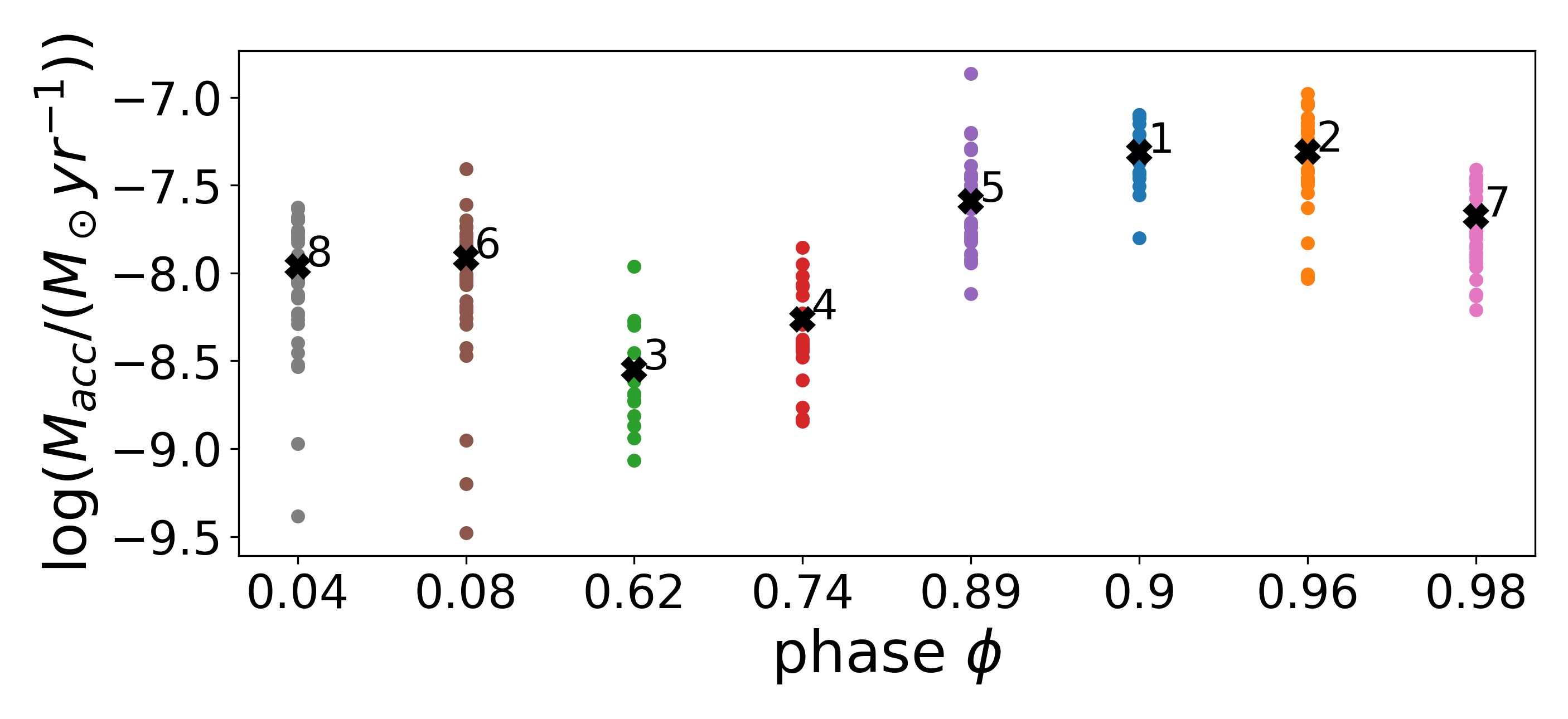}
    \end{flushright}
    \caption{{\it Top}: Filled circles shows whether the highest peak of the line corresponding to the element in y-axis in a certain epoch, has the radial velocity of the primary (red) or the secondary (blue) component, or there are two peaks of the same intensity corresponding to both the RVs (purple). Black crosses highlight line profiles for which is not possible to determine which source is accreting the most either because there is no detection, or because the two components are not spectroscopically resolved. {\it Bottom}: Co-added mass accretion rate of the two components as a function of the orbital phase, as in Fig.~\ref{fig:lacc}.}
    \label{fig:acc_summary}
\end{figure}
We identified which source is accreting the most by studying the velocity of the two peaks. We interpret that the primary (secondary) is accreting the most when the highest peak velocity corresponds to the RV of the primary (secondary). Fig.~\ref{fig:acc_summary} shows this result by putting red for primary and blue for secondary filled circle symbols. 
When the two peaks, corresponding to the two RVs, are equal, then both components likely accrete at equal rates (purple filled circle). All the other cases, when it is not possible to provide information on accretion on single sources from these lines, are marked with a cross. Thus, according to Fig.~\ref{fig:acc_summary}, the primary is accreting the most during Epoch~1, at $\phi = 0.9$, and the secondary is accreting the most during Epoch~7, at $\phi=0.98$. For all other epochs, the main accretor changes depending on which accretion tracer line is considered.

A plausible interpretation of this change in the main accretor is that different lines are formed in different regions of the accretion column.  
It has been shown that Balmer series lines are thought to be emitted by the pre-shock gas, in the outermost part of the inner disk. On the contrary, Paschen lines 
are mostly emitted in the post-shock gas \citep[][and references therein]{alc14, har16}. 
Thus, the H\,{\footnotesize I} lines we see at the same epoch are tracing the flow which is reaching the shock region (Balmer series, Fig.~\ref{fig:toff_figs}) or the gas that has already been shocked (Paschen series, Fig.~\ref{fig:toff_figsNIR}). 
In other words, Paschen 
lines show the accretion material in a region closer to the stellar surface while the material traced by the Balmer lines is still farther away from the stellar surface. Thus, looking at the same epoch but at different lines in Fig.~\ref{fig:toff_figs_text}, we see emission from the accreting flows in different regions. Because the main accretor changes with the orbital phase, sometimes we see ``new" material (Balmer series) which is going to accrete a certain component, while ``old" material (Paschen series) is accreting on the other component.

In general, narrow components (NCs) are usually formed in the post-shock region, close to the stellar surface, while broad components (BCs) in the pre-shock region \citep[][]{har16}. We should also remember that actual accretion flows are not homogeneous in density or temperature, so they could be explained as a superposition of accretion columns \citep[][]{har16}. This means that, in principle, different elements with different density and temperature develop different accretion columns. This may explain why using different lines gives different accretion rates. Averaging the different values provides the mean value of the accretion flow in the overall inner disk. Focusing only on a certain line provides a different estimate, tracing accretion from a specific region, under specific physical conditions. 

In our analysis, we prefer to analyze narrow lines because they are spectroscopically resolved and formed closer to the stellar surface. In that sense, they provide a better estimate of the amount of material that is going to fall on a certain component. For our purpose, it is also important that the line is not blue or redshifted. According to the literature \citep[][and references therein]{har16}, H$\alpha$ and [{O\,{\footnotesize I}}] lines are usually blueshifted by wind and/or jets, while H$\beta$, Pa$\gamma$, {Na\,{\footnotesize I}} doublet are usually redshifted with respect the velocity of the star, by the magnetospheric infall. Some lines might also be both blue and redshifted, this is the case of {He\,{\footnotesize I}} at $1083.0$~nm. Therefore, these lines are not suitable for our purpose. On the contrary, {He\,{\footnotesize I}} at $587.65$~nm and {Ca\,{\footnotesize II}} triplet emission lines have usually a very narrow component and are centered at the stellar velocity, likely tracing the post-shocked region, making these lines perfect for our study. Unfortunately, as shown in Fig.~\ref{fig:toff_figs_text}, the {He\,{\footnotesize I}} at $\lambda = 587.65$~nm line presents two peaks only in one epoch. For this reason, we decided that for the epochs in which the main accretor changes by looking at different lines: we established which component accretes the most by examining the {Ca\,{\footnotesize II}} triplet. For Epoch~4, where {Ca\,{\footnotesize II}} is not detected, we assumed that the secondary is the main accretor because we see differently only at H$\alpha$. The H$\alpha$ line is generally strong and broad, and being sensitive to the blue shifting due to the optically thick regime \citep[][]{har16}, is not reliable for this kind of analysis. 

Based on these arguments and looking at Fig.~\ref{fig:acc_summary}, we can conclude there is a periodic trend: before the periastron ($\phi=0.98$) the secondary is accreting the most, then, right after the periastron ($\phi = 0.04$, Epoch~8), the primary is the main accretor and in the post-shocked region, according to the Paschen series, the accretion is equally distributed between the two components, suggesting that in the future the main accretor will change. 
Indeed, the primary is still the main accretor at $\phi = 0.08$ (Epoch~6), even if, looking at Pa$\beta$ and Pa$\gamma$ the secondary is already accreting the most.  
Unfortunately, at $\phi=0.62$ (Epoch~3), the two components have the same RV, so we are not able to evaluate accretion on single stars for this epoch. 
Later, at $\phi=0.74$ (Epoch~4), the secondary is accreting the most.  
Then the accretion flows change again, directed toward the primary at $\phi = 0.89$ and 0.90, which are Epochs~5 and 1, respectively.   
Finally, at $\phi = 0.96$ (Epoch~2), the secondary is still the main accretor, while some lines predict the primary is going to be the component which will accrete the most, as it is indeed at Epoch~7 ($\phi = 0.98$). 
We note that the change in the main accretor is not related to the absolute value of the mass accretion rate.

We would like to stress that what we know about the region where a line is produced is based on single stars, whose accretion is driven by a strong magnetic field whose main component is a dipole, and the disk we refer to is the circumstellar disk. DQ~Tau is a binary system that 
can be treated as a single star, with the circumbinary disk acting as the circumstellar disk in a single star, nearby periastron, when the magnetic field of the two components merges in a dipole magnetic field \citep[][]{sal10}. 
This is all we can say based on the emission lines. 
Further insights may be gained from studying the magnetic fields that ultimately determines the path of accretion in the magnetospheric accretion model. 
Mapping the magnetic field of stars is possible, e.g., using spectropolarimetric methods, and such a study for DQ Tau is in progress using CFHT/ESPaDOnS spectra (Pouilly et al., in prep.).

The fact that the main accreting component changes is, in principle, not in agreement with numerical simulations that predict that the secondary should always be the dominant accretor in a binary system \citep[e.g.][]{hay07, hay13, far14, youcla15}. 
These predictions are based on the fact that the angular momentum for the two sources should be the same, and as the secondary is less massive, it is located farther from the center of mass of the system. 
In other words, the secondary is closer to the CBD, where it should be easier to trap the accreting flow. 
But, for DQ~Tau, because the two components have very similar masses, 
it is not surprising that the main accretor can change epoch by epoch, and this still satisfy the theoretical basis of the numerical simulations. 
However, the change in the main accreting component is predicted by both \citet[][]{gun02} and \citet[][]{mun16}, although the timescales for which such switch of the main accretor is considered is about 100 orbital period, while we see the change in the main accretor component three times in less than two orbital periods.

\section{Summary and Conclusions} \label{sect:conclusions}

We studied the accretion variability of the young eccentric binary system DQ~Tau, by analyzing 8 epochs of X-Shooter spectra taken over 7 orbital periods. We estimated the RVs, the veiling, the spectral type, the extinction and the disk mass of DQ~Tau system. We subtracted the chromospheric and the photospheric contributions from the observed spectra, to isolate the accretion component. We studied the accretion process of DQ~Tau system comparing it with single stars, previous literature about DQ~Tau, and other binary systems. We were able to select the main accretor epoch by epoch, and discuss its variability. Our main conclusions are:
\begin{itemize}
    \item DQ~Tau is a binary system composed of two early-M type stars (M0), with extinction of $A_V = 1.72~\pm~0.26$~mag, in agreement with the previous literature.
    \item We measured RVs, resolving the RVs of the two DQ~Tau components for seven epochs. Our results are in agreement with the Keplerian solutions by \citet[][]{cze16}.
    \item The veiling changes with orbital phase, time, and wavelength, in a similar way to accretion variability, confirming its correlation with the mass accretion rate.
    \item Our results on $\macc$ variability with the phase confirm the pulsed accretion model, showing larger mass accretion rate nearby periastron and are in agreement with previous results of DQ~Tau mass accretion rate computed using different techniques.
    \item The accretion luminosity and the mass accretion rate of DQ~Tau are compatible with $\lacc$ and $\macc$ of 2-3~Myr old CTTs, following the same relations with respect to the stellar parameters, if we consider the contribution of the two components.
    \item The main accretor in DQ~Tau system varies epoch by epoch. 
    This contradicts those simulations that predict that the secondary is always the main accretor in eccentric binary systems. 
    Other simulations predict that the main accretor should change, but on significantly longer time scales than what we observe here.
    
    \end{itemize}


\acknowledgments

This project has received funding from the European Research Council (ERC) under the European Union's Horizon 2020 research and innovation programme under grant agreement No 716155 (SACCRED).

This work is based on observations collected at the European Southern Observatory under ESO programme 090.D-0446 by Salter \& Kafka
We thank Dr. Stella Kafka for kindly providing us the X-Shooter date used in this work.

We thank the anonymous referee for their effort in reviewing this paper, which help us to improve it.

\vspace{5mm}
\facilities{ESO-VLT/XSHOOTER}


\software{molecfit \citep{smette2015,kausch2015}; 
{SAPHIRES {\url{https://github.com/tofflemire/saphires}}}; 
barycorrpy \citep[][]{kanodia2018}.}




\appendix

\section{Flux calibration} \label{app:flux_cal}
\begin{figure*}[t]
    \centering
    \includegraphics[width=\textwidth]{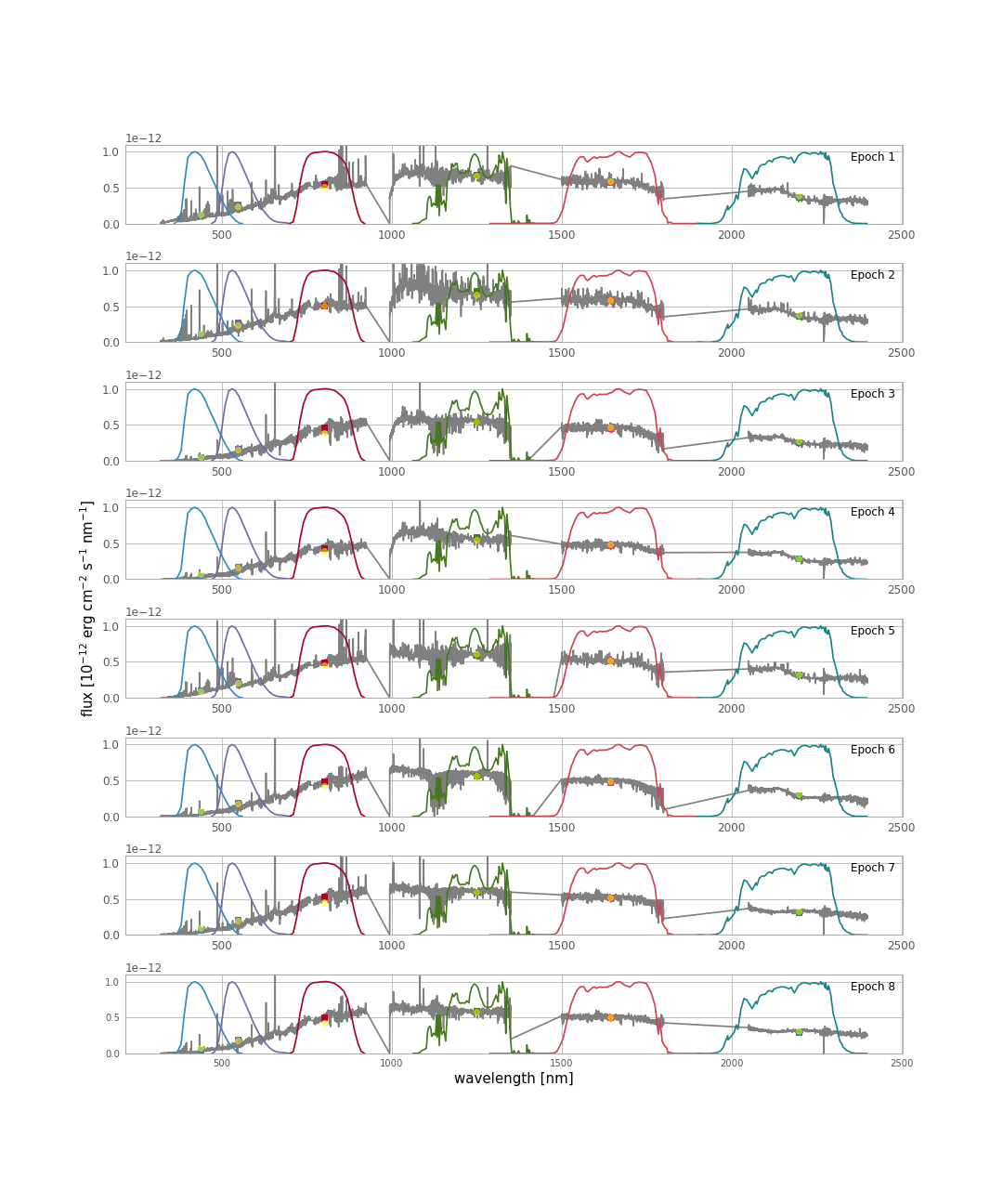}
    \caption{X-Shooter flux-calibrated spectrum from Epoch~1 (top) to Epoch~8 (bottom) are shown in grey. Yellow circles represent the observed or inferred photometry at each band. The filter bandpasses (lines) and the synthetic photometry (filled squares) obtained integrating the spectrum over these bandpasses are overlaid.}
    \label{fig:smoothed_spec_with_bandpasses}
\end{figure*}

In order to check the quality of our flux calibration, we compared the photometry we used (observed or inferred depending on the epoch) with the synthetic photometry obtained by convolving the spectrum with the filter bandpasses, band by band. 
Figure~\ref{fig:smoothed_spec_with_bandpasses} shows the general agreement within the observed/inferred and the synthetic photometry, for each epoch.

\begin{figure*}[t]
    \centering
    \includegraphics[width=\textwidth]{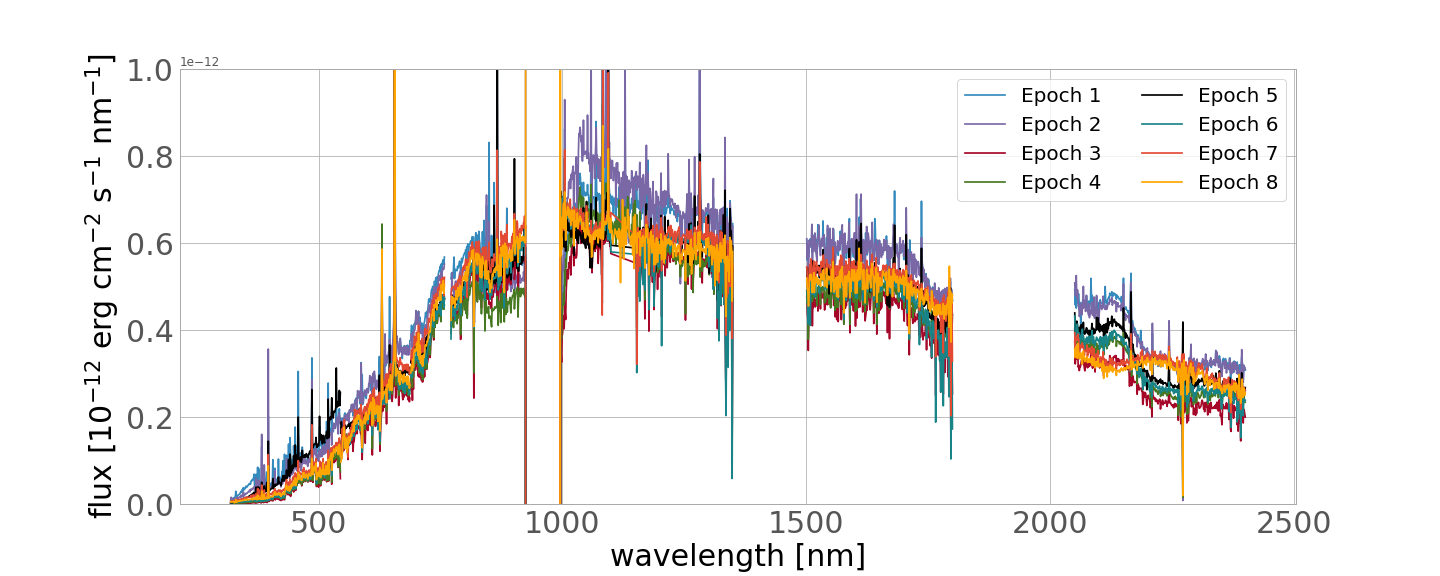}
    \caption{DQ~Tau smoothed and flux calibrated spectra.}
    \label{fig:smoothed_spec}
\end{figure*}

\section{All the accretion lines} \label{app:lines}
We present in this section all the lines profile that trace accretion in DQ~Tau. Results are summarized in the text, in Fig.~\ref{fig:acc_summary}. 

\begin{figure*}[t]
    \centering
    \includegraphics[width=0.3\textwidth]{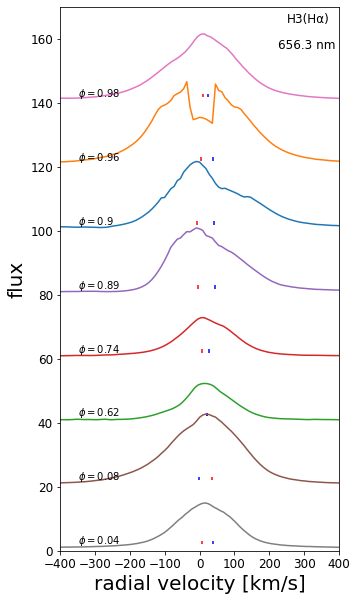}
    \includegraphics[width=0.3\textwidth]{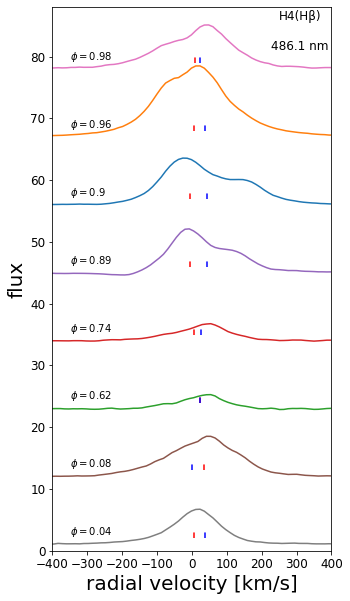}
    \includegraphics[width=0.3\textwidth]{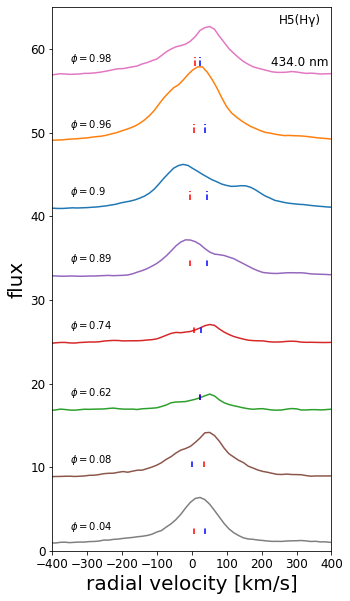}
    \includegraphics[width=0.3\textwidth]{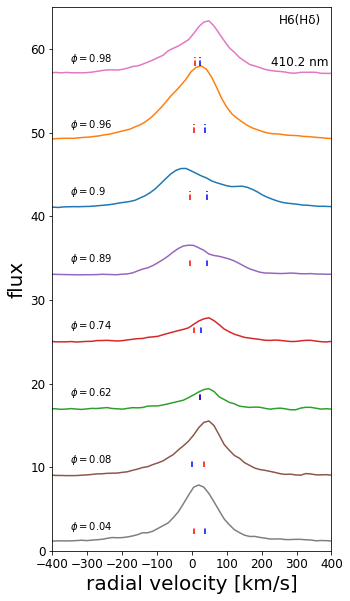}
    \includegraphics[width=0.3\textwidth]{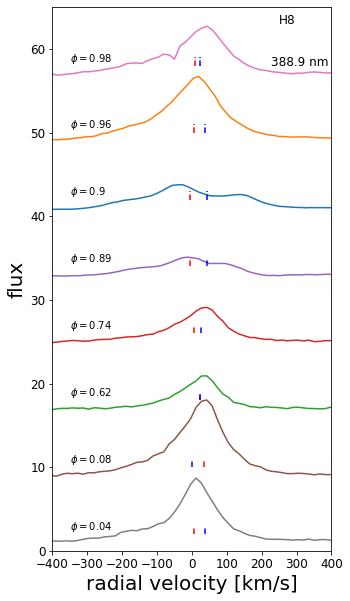}
    \includegraphics[width=0.3\textwidth]{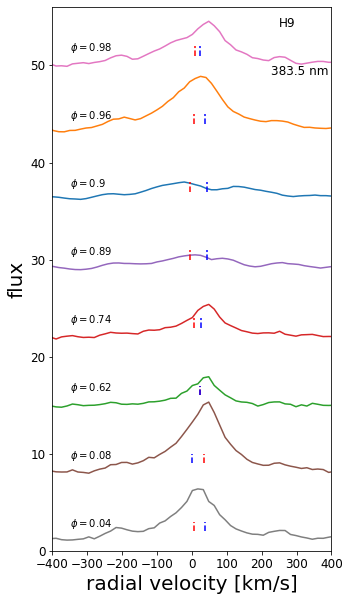}
    \caption{Velocity structure of continuum-normalized {H\,{\footnotesize I}} emission lines as indicated in the figures, observed by X-Shooter in UVB and VIS. Vertical red and blue dashed lines are velocities of the two primary and secondary component, respectively. Spectra are ordered from bottom to top by increasing orbital phase, which is labeled adjacent to
each spectrum. The line color marks the epoch, using the same colors of Fig.~\ref{fig:lacc}.}
    \label{fig:toff_figs}
\end{figure*}
\begin{figure*}[t]
    \centering
    \includegraphics[width=0.3\textwidth]{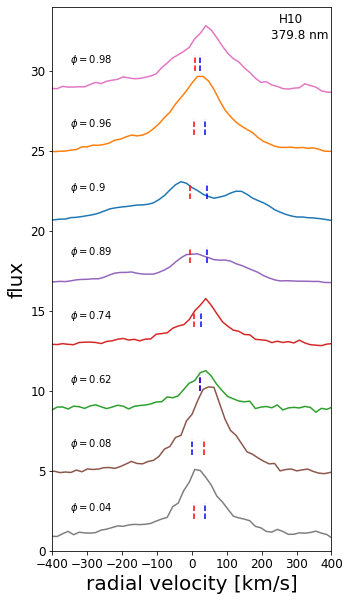}
    \includegraphics[width=0.3\textwidth]{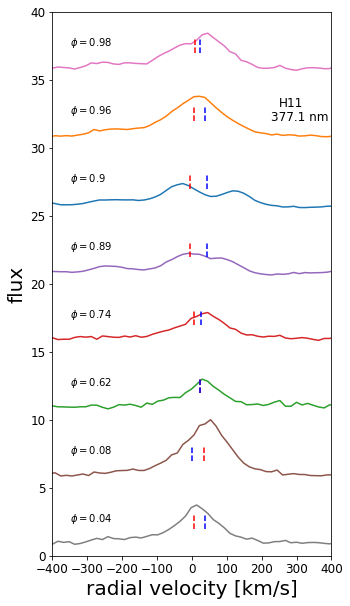}
    \includegraphics[width=0.3\textwidth]{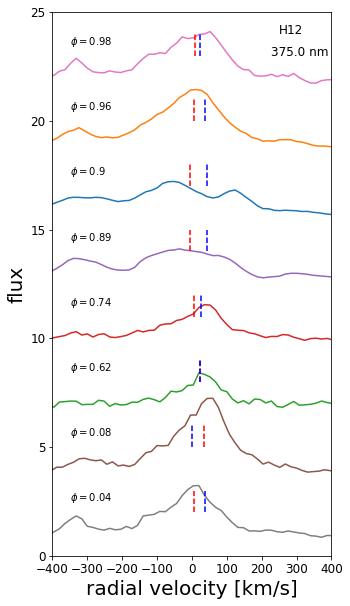}
    \includegraphics[width=0.3\textwidth]{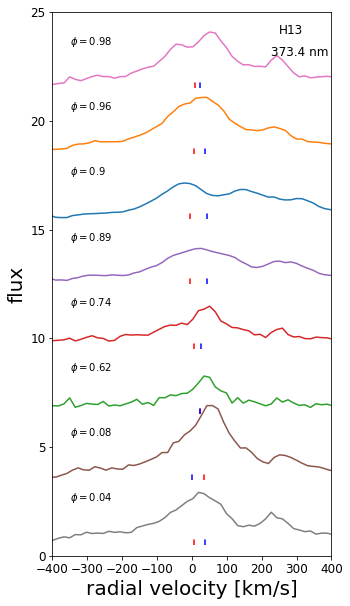}
    \includegraphics[width=0.3\textwidth]{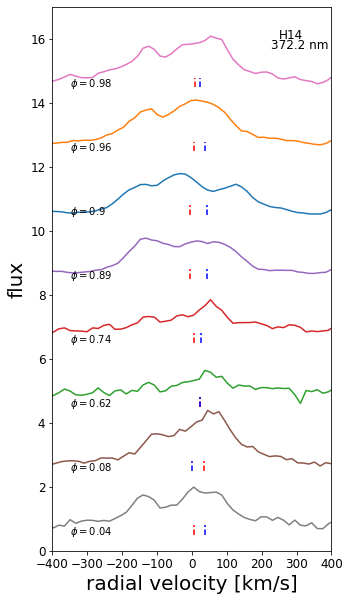}
    \includegraphics[width=0.3\textwidth]{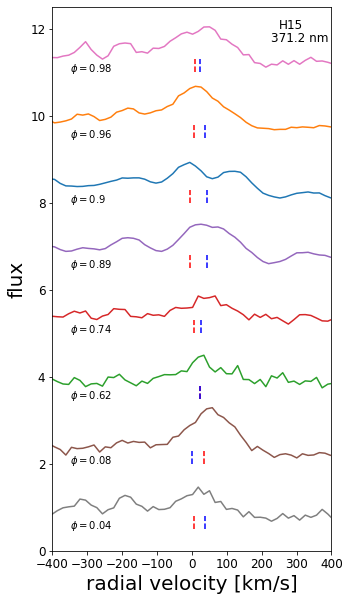}
    \caption{Velocity structure of continuum-normalized {H\,{\footnotesize I}} emission lines as indicated in the figures, observed by X-Shooter in UVB and VIS. Vertical red and blue dashed lines are velocities of the primary and secondary component, respectively. Spectra are ordered from bottom to top by increasing orbital phase, which is labeled adjacent to
each spectrum. The line color marks the epoch, using the same colors of Fig.~\ref{fig:lacc}.}
    \label{fig:toff_figs1}
\end{figure*}

\begin{figure*}[t]
    \centering
    \includegraphics[width=0.3\textwidth]{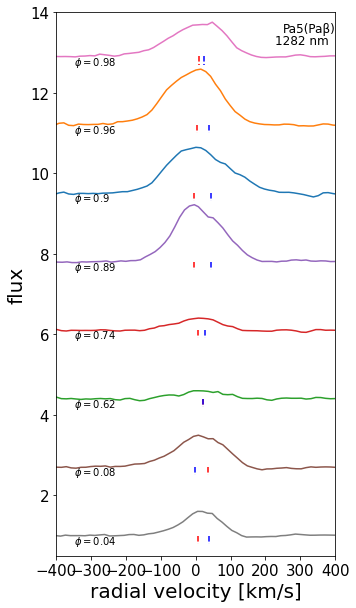}
    \includegraphics[width=0.3\textwidth]{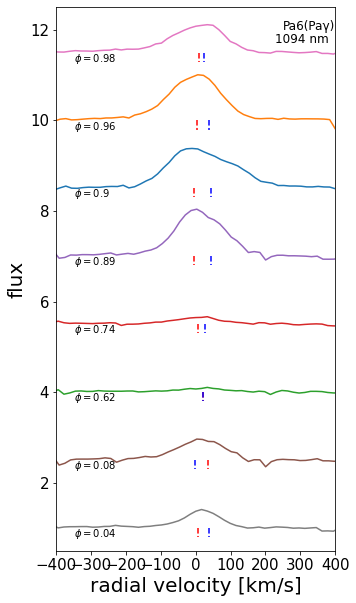}
    \includegraphics[width=0.3\textwidth]{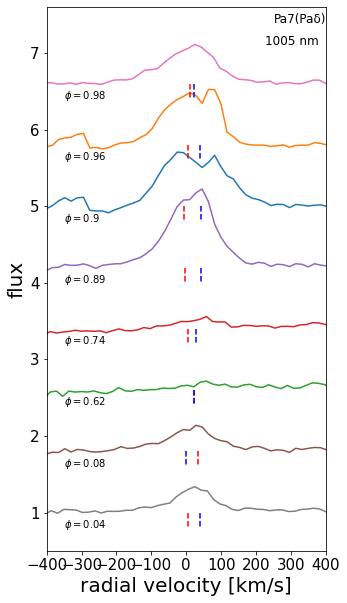}
    \includegraphics[width=0.3\textwidth]{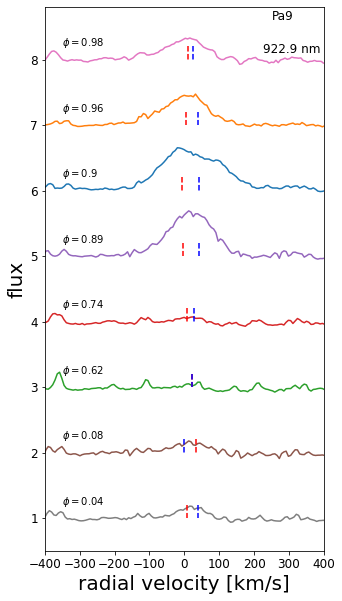}
    \includegraphics[width=0.3\textwidth]{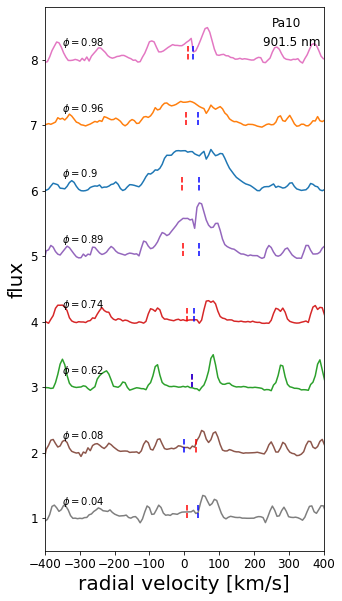}
    \includegraphics[width=0.3\textwidth]{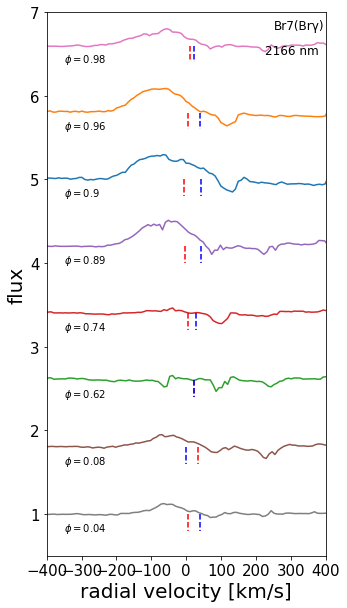}
\caption{Velocity structure of continuum-normalized {H\,{\footnotesize I}} emission lines as indicated in the figures, observed by X-Shooter in NIR. Vertical red and blue dashed lines are velocities of the primary and secondary component, respectively. Spectra are ordered from bottom to top by increasing orbital phase, which is labeled adjacent to
each spectrum. The line color marks the epoch, using the same colors of Fig.~\ref{fig:lacc}.}
    \label{fig:toff_figsNIR}
\end{figure*}

\begin{figure*}[t]
    \centering
    \includegraphics[width=0.3\textwidth]{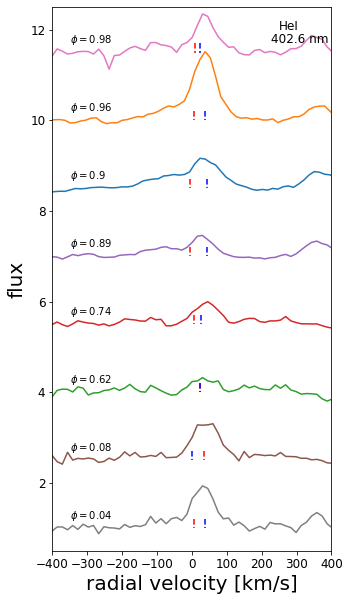}
    \includegraphics[width=0.3\textwidth]{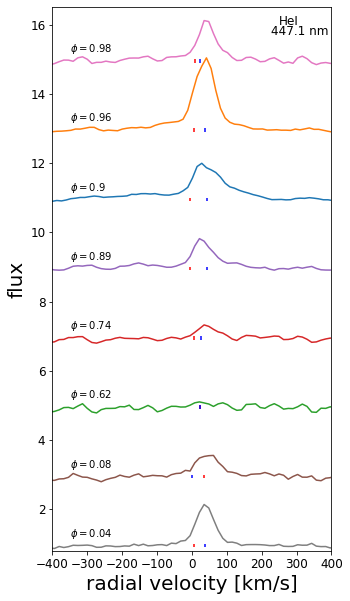}
    \includegraphics[width=0.3\textwidth]{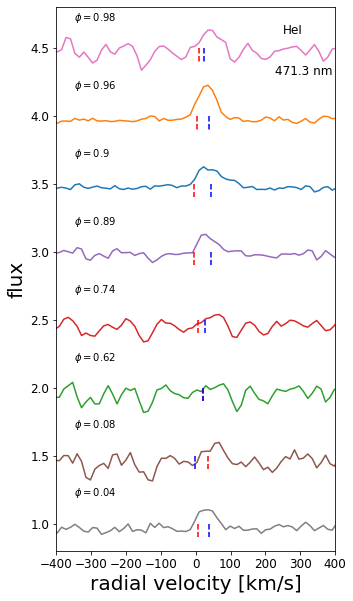}
    \includegraphics[width=0.3\textwidth]{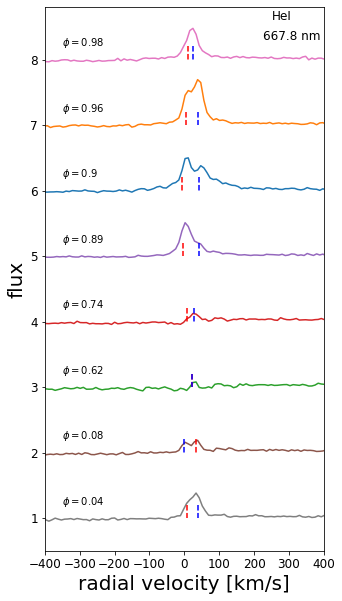}
    \includegraphics[width=0.3\textwidth]{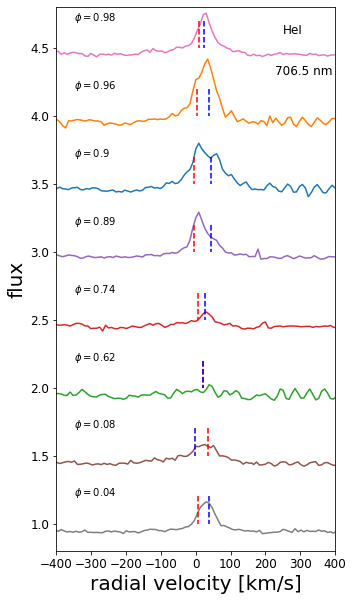}
    \includegraphics[width=0.3\textwidth]{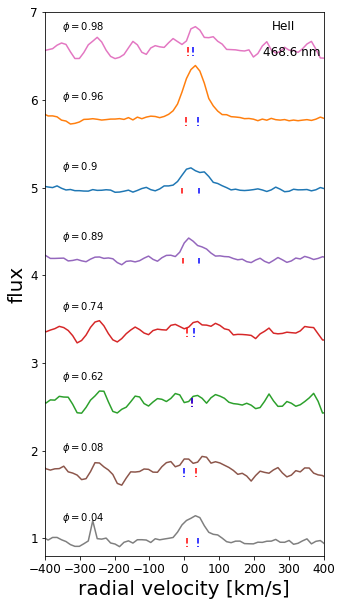}
    \caption{Velocity structure of continuum-normalized {He\,{\footnotesize I}} and {He\,{\footnotesize II}} emission lines as indicated in the figures, observed by X-Shooter. Vertical red and blue dashed lines are velocities of the primary and secondary component, respectively. Spectra are ordered from bottom to top by increasing orbital phase, which is labeled adjacent to
each spectrum. The line color marks the epoch, using the same colors of Fig.~\ref{fig:lacc}.}
    \label{fig:toff_figsHe}
\end{figure*}

\begin{figure*}[t]
    \centering
    \includegraphics[width=0.3\textwidth]{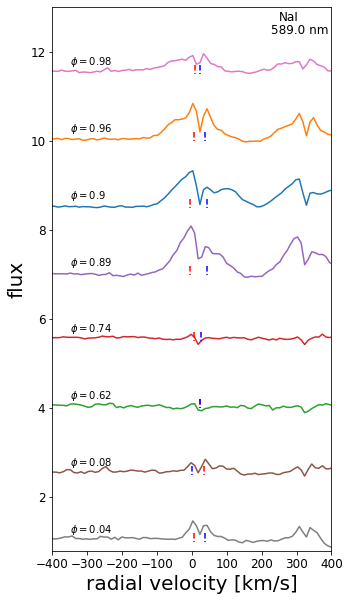}
    \includegraphics[width=0.3\textwidth]{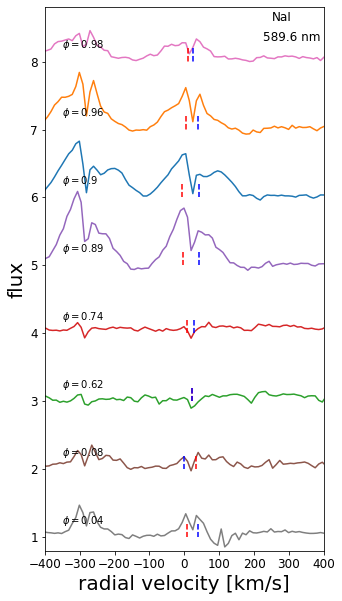}
    \includegraphics[width=0.3\textwidth]{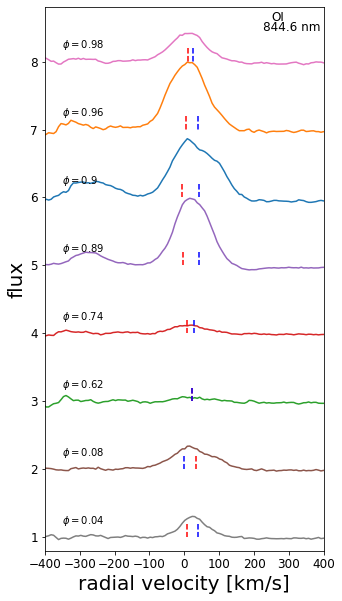}
    \caption{Velocity structure of continuum-normalized {Ca\,{\footnotesize II}} triplet, {Na\,{\footnotesize I}} doublet, and {O\,{\footnotesize I}} emission lines as indicated in the figures, observed by X-Shooter. Vertical red and blue dashed lines are velocities of the primary and secondary component, respectively. Spectra are ordered from bottom to top by increasing orbital phase, which is labeled adjacent to
each spectrum. The line color marks the epoch, using the same colors of Fig.~\ref{fig:lacc}.}
    \label{fig:toff_figs_other}
\end{figure*}


\bibliography{biblio.bib}{}

\begin{thebibliography}{}
\expandafter\ifx\csname natexlab\endcsname\relax\def\natexlab#1{#1}\fi
\providecommand{\url}[1]{\href{#1}{#1}}
\providecommand{\dodoi}[1]{doi:~\href{http://doi.org/#1}{\nolinkurl{#1}}}
\providecommand{\doeprint}[1]{\href{http://ascl.net/#1}{\nolinkurl{http://ascl.net/#1}}}
\providecommand{\doarXiv}[1]{\href{https://arxiv.org/abs/#1}{\nolinkurl{https://arxiv.org/abs/#1}}}

\bibitem[{{Alcal{\'a}} {et~al.}(2019){Alcal{\'a}}, {Manara}, {France},
  {Schneider}, {Arulanantham}, {Miotello}, {G{\"u}nther}, \& {Brown}}]{alc19}
{Alcal{\'a}}, J.~M., {Manara}, C.~F., {France}, K., {et~al.} 2019, \aap, 629,
  A108, \dodoi{10.1051/0004-6361/201935657}

\bibitem[{{Alcal{\'a}} {et~al.}(2014){Alcal{\'a}}, {Natta}, {Manara}, {Spezzi},
  {Stelzer}, {Frasca}, {Biazzo}, {Covino}, {Randich}, {Rigliaco}, {Testi},
  {Comer{\'o}n}, {Cupani}, \& {D'Elia}}]{alc14}
{Alcal{\'a}}, J.~M., {Natta}, A., {Manara}, C.~F., {et~al.} 2014, \aap, 561,
  A2, \dodoi{10.1051/0004-6361/201322254}

\bibitem[{{Alcal{\'a}} {et~al.}(2017){Alcal{\'a}}, {Manara}, {Natta}, {Frasca},
  {Testi}, {Nisini}, {Stelzer}, {Williams}, {Antoniucci}, \& {Biazzo}}]{alc17}
{Alcal{\'a}}, J.~M., {Manara}, C.~F., {Natta}, A., {et~al.} 2017, \aap, 600,
  A20, \dodoi{10.1051/0004-6361/201629929}

\bibitem[{{Alcal{\'a}} {et~al.}(2021){Alcal{\'a}}, {Gangi}, {Biazzo},
  {Antoniucci}, {Frasca}, {Giannini}, {Munari}, {Nisini}, {Harutyunyan},
  {Manara}, \& {Vitali}}]{alc21}
{Alcal{\'a}}, J.~M., {Gangi}, M., {Biazzo}, K., {et~al.} 2021, \aap, 652, A72,
  \dodoi{10.1051/0004-6361/202140918}

\bibitem[{{Andrews} {et~al.}(2011){Andrews}, {Wilner}, {Espaillat}, {Hughes},
  {Dullemond}, {McClure}, {Qi}, \& {Brown}}]{andrews11}
{Andrews}, S.~M., {Wilner}, D.~J., {Espaillat}, C., {et~al.} 2011, \apj, 732,
  42, \dodoi{10.1088/0004-637X/732/1/42}

\bibitem[{{Antoniucci} {et~al.}(2014){Antoniucci}, {Giannini}, {Li Causi}, \&
  {Lorenzetti}}]{ant14}
{Antoniucci}, S., {Giannini}, T., {Li Causi}, G., \& {Lorenzetti}, D. 2014,
  \apj, 782, 51, \dodoi{10.1088/0004-637X/782/1/51}

\bibitem[{{Artymowicz} \& {Lubow}(1994)}]{art94}
{Artymowicz}, P., \& {Lubow}, S.~H. 1994, \apj, 421, 651,
  \dodoi{10.1086/173679}

\bibitem[{{Artymowicz} \& {Lubow}(1996)}]{art96}
---. 1996, \apjl, 467, L77, \dodoi{10.1086/310200}

\bibitem[{{Bailer-Jones} {et~al.}(2021){Bailer-Jones}, {Rybizki}, {Fouesneau},
  {Demleitner}, \& {Andrae}}]{bailerjones21}
{Bailer-Jones}, C.~A.~L., {Rybizki}, J., {Fouesneau}, M., {Demleitner}, M., \&
  {Andrae}, R. 2021, \aj, 161, 147, \dodoi{10.3847/1538-3881/abd806}

\bibitem[{{Basri} \& {Batalha}(1990)}]{basri1990}
{Basri}, G., \& {Batalha}, C. 1990, \apj, 363, 654, \dodoi{10.1086/169374}

\bibitem[{{Cardelli} {et~al.}(1989){Cardelli}, {Clayton}, \& {Mathis}}]{car89}
{Cardelli}, J.~A., {Clayton}, G.~C., \& {Mathis}, J.~S. 1989, \apj, 345, 245,
  \dodoi{10.1086/167900}

\bibitem[{{Coelho} {et~al.}(2005){Coelho}, {Barbuy}, {Mel{\'e}ndez},
  {Schiavon}, \& {Castilho}}]{coelho2005}
{Coelho}, P., {Barbuy}, B., {Mel{\'e}ndez}, J., {Schiavon}, R.~P., \&
  {Castilho}, B.~V. 2005, \aap, 443, 735, \dodoi{10.1051/0004-6361:20053511}

\bibitem[{{Czekala} {et~al.}(2016){Czekala}, {Andrews}, {Torres}, {Jensen},
  {Stassun}, {Wilner}, \& {Latham}}]{cze16}
{Czekala}, I., {Andrews}, S.~M., {Torres}, G., {et~al.} 2016, \apj, 818, 156,
  \dodoi{10.3847/0004-637X/818/2/156}

\bibitem[{{Dodin} \& {Lamzin}(2012)}]{dodin2012}
{Dodin}, A.~V., \& {Lamzin}, S.~A. 2012, Astronomy Letters, 38, 649,
  \dodoi{10.1134/S1063773712100027}

\bibitem[{{D'Orazio} {et~al.}(2013){D'Orazio}, {Haiman}, \&
  {MacFadyen}}]{dorazio13}
{D'Orazio}, D.~J., {Haiman}, Z., \& {MacFadyen}, A. 2013, \mnras, 436, 2997,
  \dodoi{10.1093/mnras/stt1787}

\bibitem[{{Erlick} {et~al.}(1998){Erlick}, {Frederick}, {Saxena}, \&
  {Wenny}}]{erl98}
{Erlick}, C., {Frederick}, J.~E., {Saxena}, V.~K., \& {Wenny}, B.~N. 1998,
  \jgr, 103, 31,541, \dodoi{10.1029/1998JD200053}

\bibitem[{{Farris} {et~al.}(2014{\natexlab{a}}){Farris}, {Duffell},
  {MacFadyen}, \& {Haiman}}]{farris14}
{Farris}, B.~D., {Duffell}, P., {MacFadyen}, A.~I., \& {Haiman}, Z.
  2014{\natexlab{a}}, \apj, 783, 134, \dodoi{10.1088/0004-637X/783/2/134}

\bibitem[{{Farris} {et~al.}(2014{\natexlab{b}}){Farris}, {Duffell},
  {MacFadyen}, \& {Haiman}}]{far14}
---. 2014{\natexlab{b}}, \apj, 783, 134, \dodoi{10.1088/0004-637X/783/2/134}

\bibitem[{{Fiorellino} {et~al.}(2021){Fiorellino}, {Manara}, {Nisini},
  {Ramsay}, {Antoniucci}, {Giannini}, {Biazzo}, {Alcal{\`a}}, \&
  {Fedele}}]{fio21}
{Fiorellino}, E., {Manara}, C.~F., {Nisini}, B., {et~al.} 2021, \aap, 650, A43,
  \dodoi{10.1051/0004-6361/202039264}

\bibitem[{{Folha} \& {Emerson}(1999)}]{folha1999}
{Folha}, D.~F.~M., \& {Emerson}, J.~P. 1999, \aap, 352, 517

\bibitem[{{Gullbring} {et~al.}(1998){Gullbring}, {Hartmann}, {Brice{\~n}o}, \&
  {Calvet}}]{gul98}
{Gullbring}, E., {Hartmann}, L., {Brice{\~n}o}, C., \& {Calvet}, N. 1998, \apj,
  492, 323, \dodoi{10.1086/305032}

\bibitem[{{G{\"u}nther} \& {Kley}(2002{\natexlab{a}})}]{gun02}
{G{\"u}nther}, R., \& {Kley}, W. 2002{\natexlab{a}}, \aap, 387, 550,
  \dodoi{10.1051/0004-6361:20020407}

\bibitem[{{G{\"u}nther} \& {Kley}(2002{\natexlab{b}})}]{gut02}
---. 2002{\natexlab{b}}, \aap, 387, 550, \dodoi{10.1051/0004-6361:20020407}

\bibitem[{{Hartigan} {et~al.}(1995){Hartigan}, {Edwards}, \&
  {Ghandour}}]{har95}
{Hartigan}, P., {Edwards}, S., \& {Ghandour}, L. 1995, \apj, 452, 736,
  \dodoi{10.1086/176344}

\bibitem[{{Hartmann} {et~al.}(1998){Hartmann}, {Calvet}, {Gullbring}, \&
  {D'Alessio}}]{har98}
{Hartmann}, L., {Calvet}, N., {Gullbring}, E., \& {D'Alessio}, P. 1998, \apj,
  495, 385, \dodoi{10.1086/305277}

\bibitem[{{Hartmann} {et~al.}(2016){Hartmann}, {Herczeg}, \& {Calvet}}]{har16}
{Hartmann}, L., {Herczeg}, G., \& {Calvet}, N. 2016, \araa, 54, 135,
  \dodoi{10.1146/annurev-astro-081915-023347}

\bibitem[{{Hayasaki} {et~al.}(2007){Hayasaki}, {Mineshige}, \& {Sudou}}]{hay07}
{Hayasaki}, K., {Mineshige}, S., \& {Sudou}, H. 2007, \pasj, 59, 427,
  \dodoi{10.1093/pasj/59.2.427}

\bibitem[{{Hayasaki} {et~al.}(2013){Hayasaki}, {Saito}, \& {Mineshige}}]{hay13}
{Hayasaki}, K., {Saito}, H., \& {Mineshige}, S. 2013, \pasj, 65, 86,
  \dodoi{10.1093/pasj/65.4.86}

\bibitem[{{Herbig}(1977)}]{her77}
{Herbig}, G.~H. 1977, \apj, 214, 747, \dodoi{10.1086/155304}

\bibitem[{{Herczeg} \& {Hillenbrand}(2014)}]{her14}
{Herczeg}, G.~J., \& {Hillenbrand}, L.~A. 2014, \apj, 786, 97,
  \dodoi{10.1088/0004-637X/786/2/97}

\bibitem[{{Kanodia} \& {Wright}(2018)}]{kanodia2018}
{Kanodia}, S., \& {Wright}, J. 2018, Research Notes of the American
  Astronomical Society, 2, 4, \dodoi{10.3847/2515-5172/aaa4b7}

\bibitem[{{Kausch} {et~al.}(2015{\natexlab{a}}){Kausch}, {Noll}, {Smette},
  {Kimeswenger}, {Barden}, {Szyszka}, {Jones}, {Sana}, {Horst}, \&
  {Kerber}}]{kau15}
{Kausch}, W., {Noll}, S., {Smette}, A., {et~al.} 2015{\natexlab{a}}, \aap, 576,
  A78, \dodoi{10.1051/0004-6361/201423909}

\bibitem[{{Kausch} {et~al.}(2015{\natexlab{b}}){Kausch}, {Noll}, {Smette},
  {Kimeswenger}, {Barden}, {Szyszka}, {Jones}, {Sana}, {Horst}, \&
  {Kerber}}]{kausch2015}
---. 2015{\natexlab{b}}, \aap, 576, A78, \dodoi{10.1051/0004-6361/201423909}

\bibitem[{{Kochanek} {et~al.}(2017){Kochanek}, {Shappee}, {Stanek}, {Holoien},
  {Thompson}, {Prieto}, {Dong}, {Shields}, {Will}, {Britt}, {Perzanowski}, \&
  {Pojma{\'n}ski}}]{koc17}
{Kochanek}, C.~S., {Shappee}, B.~J., {Stanek}, K.~Z., {et~al.} 2017, \pasp,
  129, 104502, \dodoi{10.1088/1538-3873/aa80d9}

\bibitem[{{Koornneef}(1983)}]{koo83}
{Koornneef}, J. 1983, \aap, 500, 247

\bibitem[{{K{\'o}sp{\'a}l} {et~al.}(2018){K{\'o}sp{\'a}l}, {{\'A}brah{\'a}m},
  {Zsidi}, {Vida}, {Szab{\'o}}, {Mo{\'o}r}, \& {P{\'a}l}}]{kos18}
{K{\'o}sp{\'a}l}, {\'A}., {{\'A}brah{\'a}m}, P., {Zsidi}, G., {et~al.} 2018,
  \apj, 862, 44, \dodoi{10.3847/1538-4357/aacafa}

\bibitem[{{Kurosawa} \& {Romanova}(2013)}]{kurosawa2013}
{Kurosawa}, R., \& {Romanova}, M.~M. 2013, \mnras, 431, 2673,
  \dodoi{10.1093/mnras/stt365}

\bibitem[{{Kurosawa} {et~al.}(2008){Kurosawa}, {Romanova}, \&
  {Harries}}]{kurosawa2008}
{Kurosawa}, R., {Romanova}, M.~M., \& {Harries}, T.~J. 2008, \mnras, 385, 1931,
  \dodoi{10.1111/j.1365-2966.2008.13055.x}

\bibitem[{{Manara} {et~al.}(2019){Manara}, {Mordasini}, {Testi}, {Williams},
  {Miotello}, {Lodato}, \& {Emsenhuber}}]{man19}
{Manara}, C.~F., {Mordasini}, C., {Testi}, L., {et~al.} 2019, \aap, 631, L2,
  \dodoi{10.1051/0004-6361/201936488}

\bibitem[{{Manara} {et~al.}(2013){Manara}, {Testi}, {Rigliaco}, {Alcal{\'a}},
  {Natta}, {Stelzer}, {Biazzo}, {Covino}, {Covino}, {Cupani}, {D'Elia}, \&
  {Rand ich}}]{man13a}
{Manara}, C.~F., {Testi}, L., {Rigliaco}, E., {et~al.} 2013, \aap, 551, A107,
  \dodoi{10.1051/0004-6361/201220921}

\bibitem[{{Manara} {et~al.}(2017){Manara}, {Testi}, {Herczeg}, {Pascucci},
  {Alcal{\'a}}, {Natta}, {Antoniucci}, {Fedele}, {Mulders}, {Henning},
  {Mohanty}, {Prusti}, \& {Rigliaco}}]{man17cha}
{Manara}, C.~F., {Testi}, L., {Herczeg}, G.~J., {et~al.} 2017, \aap, 604, A127,
  \dodoi{10.1051/0004-6361/201630147}

\bibitem[{{Mathieu} {et~al.}(1997){Mathieu}, {Stassun}, {Basri}, {Jensen},
  {Johns-Krull}, {Valenti}, \& {Hartmann}}]{mathieu1997}
{Mathieu}, R.~D., {Stassun}, K., {Basri}, G., {et~al.} 1997, \aj, 113, 1841,
  \dodoi{10.1086/118395}

\bibitem[{{Meyer} {et~al.}(1997){Meyer}, {Calvet}, \& {Hillenbrand}}]{mey97}
{Meyer}, M.~R., {Calvet}, N., \& {Hillenbrand}, L.~A. 1997, \aj, 114, 288,
  \dodoi{10.1086/118474}

\bibitem[{{Monin} {et~al.}(2007){Monin}, {Clarke}, {Prato}, \&
  {McCabe}}]{mon07}
{Monin}, J.~L., {Clarke}, C.~J., {Prato}, L., \& {McCabe}, C. 2007, in
  Protostars and Planets V, ed. B.~{Reipurth}, D.~{Jewitt}, \& K.~{Keil}, 395

\bibitem[{{Mu{\~n}oz} \& {Lai}(2016)}]{mun16}
{Mu{\~n}oz}, D.~J., \& {Lai}, D. 2016, \apj, 827, 43,
  \dodoi{10.3847/0004-637X/827/1/43}

\bibitem[{{Muzerolle} {et~al.}(2019){Muzerolle}, {Flaherty}, {Balog}, {Beck},
  \& {Gutermuth}}]{muz19}
{Muzerolle}, J., {Flaherty}, K., {Balog}, Z., {Beck}, T., \& {Gutermuth}, R.
  2019, \apj, 877, 29, \dodoi{10.3847/1538-4357/ab1756}

\bibitem[{{Muzerolle} {et~al.}(1998){Muzerolle}, {Hartmann}, \&
  {Calvet}}]{muz98tau}
{Muzerolle}, J., {Hartmann}, L., \& {Calvet}, N. 1998, \aj, 116, 455,
  \dodoi{10.1086/300428}

\bibitem[{{Newnham} \& {Ballard}(1998)}]{new98}
{Newnham}, D.~A., \& {Ballard}, J. 1998, \jgr, 103, 28,801,
  \dodoi{10.1029/98JD02799}

\bibitem[{{Rei} {et~al.}(2018){Rei}, {Petrov}, \& {Gameiro}}]{rei2018}
{Rei}, A.~C.~S., {Petrov}, P.~P., \& {Gameiro}, J.~F. 2018, \aap, 610, A40,
  \dodoi{10.1051/0004-6361/201731444}

\bibitem[{{Rigliaco} {et~al.}(2012){Rigliaco}, {Natta}, {Testi}, {Randich},
  {Alcal{\`a}}, {Covino}, \& {Stelzer}}]{rig12}
{Rigliaco}, E., {Natta}, A., {Testi}, L., {et~al.} 2012, \aap, 548, A56,
  \dodoi{10.1051/0004-6361/201219832}

\bibitem[{{Romanova} \& {Owocki}(2016)}]{romanova2016}
{Romanova}, M.~M., \& {Owocki}, S.~P. 2016, {Accretion, Outflows, and Winds of
  Magnetized Stars}, Vol.~54, 347

\bibitem[{{Salter} {et~al.}(2010){Salter}, {K{\'o}sp{\'a}l}, {Getman},
  {Hogerheijde}, {van Kempen}, {Carpenter}, {Blake}, \& {Wilner}}]{sal10}
{Salter}, D.~M., {K{\'o}sp{\'a}l}, {\'A}., {Getman}, K.~V., {et~al.} 2010,
  \aap, 521, A32, \dodoi{10.1051/0004-6361/201015197}

\bibitem[{{Shappee} {et~al.}(2014){Shappee}, {Prieto}, {Grupe}, {Kochanek},
  {Stanek}, {De Rosa}, {Mathur}, {Zu}, {Peterson}, {Pogge}, {Komossa}, {Im},
  {Jencson}, {Holoien}, {Basu}, {Beacom}, {Szczygie{\l}}, {Brimacombe},
  {Adams}, {Campillay}, {Choi}, {Contreras}, {Dietrich}, {Dubberley},
  {Elphick}, {Foale}, {Giustini}, {Gonzalez}, {Hawkins}, {Howell}, {Hsiao},
  {Koss}, {Leighly}, {Morrell}, {Mudd}, {Mullins}, {Nugent}, {Parrent},
  {Phillips}, {Pojmanski}, {Rosing}, {Ross}, {Sand}, {Terndrup}, {Valenti},
  {Walker}, \& {Yoon}}]{sha14}
{Shappee}, B.~J., {Prieto}, J.~L., {Grupe}, D., {et~al.} 2014, \apj, 788, 48,
  \dodoi{10.1088/0004-637X/788/1/48}

\bibitem[{{Smette} {et~al.}(2015){Smette}, {Sana}, {Noll}, {Horst}, {Kausch},
  {Kimeswenger}, {Barden}, {Szyszka}, {Jones}, {Gallenne}, {Vinther},
  {Ballester}, \& {Taylor}}]{smette2015}
{Smette}, A., {Sana}, H., {Noll}, S., {et~al.} 2015, \aap, 576, A77,
  \dodoi{10.1051/0004-6361/201423932}

\bibitem[{{Tofflemire} {et~al.}(2017){Tofflemire}, {Mathieu}, {Ardila},
  {Akeson}, {Ciardi}, {Johns-Krull}, {Herczeg}, \& {Quijano-Vodniza}}]{tof17}
{Tofflemire}, B.~M., {Mathieu}, R.~D., {Ardila}, D.~R., {et~al.} 2017, \apj,
  835, 8, \dodoi{10.3847/1538-4357/835/1/8}

\bibitem[{{Tofflemire} {et~al.}(2019){Tofflemire}, {Mathieu}, \&
  {Johns-Krull}}]{tof19}
{Tofflemire}, B.~M., {Mathieu}, R.~D., \& {Johns-Krull}, C.~M. 2019, \aj, 158,
  245, \dodoi{10.3847/1538-3881/ab4f7d}

\bibitem[{{Vernet} {et~al.}(2011){Vernet}, {Dekker}, {D'Odorico}, {Kaper},
  {Kjaergaard}, {Hammer}, {Randich}, {Zerbi}, {Groot}, {Hjorth}, {Guinouard},
  {Navarro}, {Adolfse}, {Albers}, {Amans}, {Andersen}, {Andersen}, {Binetruy},
  {Bristow}, {Castillo}, {Chemla}, {Christensen}, {Conconi}, {Conzelmann},
  {Dam}, {de Caprio}, {de Ugarte Postigo}, {Delabre}, {di Marcantonio},
  {Downing}, {Elswijk}, {Finger}, {Fischer}, {Flores}, {Fran{\c{c}}ois},
  {Goldoni}, {Guglielmi}, {Haigron}, {Hanenburg}, {Hendriks}, {Horrobin},
  {Horville}, {Jessen}, {Kerber}, {Kern}, {Kiekebusch}, {Kleszcz}, {Klougart},
  {Kragt}, {Larsen}, {Lizon}, {Lucuix}, {Mainieri}, {Manuputy}, {Martayan},
  {Mason}, {Mazzoleni}, {Michaelsen}, {Modigliani}, {Moehler}, {M{\o}ller},
  {Norup S{\o}rensen}, {N{\o}rregaard}, {P{\'e}roux}, {Patat}, {Pena}, {Pragt},
  {Reinero}, {Rigal}, {Riva}, {Roelfsema}, {Royer}, {Sacco}, {Santin},
  {Schoenmaker}, {Spano}, {Sweers}, {Ter Horst}, {Tintori}, {Tromp}, {van
  Dael}, {van der Vliet}, {Venema}, {Vidali}, {Vinther}, {Vola}, {Winters},
  {Wistisen}, {Wulterkens}, \& {Zacchei}}]{ver11}
{Vernet}, J., {Dekker}, H., {D'Odorico}, S., {et~al.} 2011, \aap, 536, A105,
  \dodoi{10.1051/0004-6361/201117752}

\bibitem[{{Young} \& {Clarke}(2015)}]{youcla15}
{Young}, M.~D., \& {Clarke}, C.~J. 2015, \mnras, 452, 3085,
  \dodoi{10.1093/mnras/stv1512}

\end{thebibliography}
\bibliographystyle{aasjournal}



\end{document}